\documentclass[lettersize,journal]{IEEEtran}
\usepackage{amsmath,amsfonts}
\usepackage{array}
\usepackage{textcomp}
\usepackage{stfloats}
\usepackage{url}
\usepackage{verbatim}
\usepackage{graphicx}
\usepackage{cite}
\usepackage{color}
\usepackage{mathtools}
\usepackage[version=4]{mhchem}
\usepackage[per-mode=symbol]{siunitx}
\usepackage{tabularx,booktabs}
\usepackage{xurl}
\usepackage[ruled]{algorithm2e}
\usepackage{physics}
\usepackage{here}
\usepackage{optidef}
\usepackage{subcaption}
\usepackage{amsmath,amssymb,amsthm}
\usepackage{cleveref}

\crefrangelabelformat{section}{#3#1#4--#5#2#6}

\SetArgSty{textnormal}

\newif\ifblind
\blindfalse  
\newcommand{\blind}[2]{\ifblind#1\else#2\fi}

\newif\ifpreprint
\preprinttrue  
\newcommand{\preprint}[2]{\ifpreprint#1\else#2\fi}

\newlength{\figwidth}             
\newcommand{\calF}{{\mathcal F}}
\newcommand{\calN}{{\mathcal N}}
\newcommand{\calU}{{\mathcal U}}

\newcommand{\bbE}{{\mathbb E}}
\newcommand{\bbN}{{\mathbb N}}
\newcommand{\bbR}{{\mathbb R}}

\newcommand{\bfe}{{\mathbf e}}
\newcommand{\bfu}{{\mathbf u}}
\newcommand{\bfv}{{\mathbf v}}
\newcommand{\bfw}{{\mathbf w}}
\newcommand{\bfx}{{\mathbf x}}
\newcommand{\bfy}{{\mathbf y}}
\newcommand{\bfz}{{\mathbf z}}

\newcommand{\bfA}{{\mathbf A}}
\newcommand{\bfB}{{\mathbf B}}
\newcommand{\bfF}{{\mathbf F}}
\newcommand{\bfG}{{\mathbf G}}
\newcommand{\bfP}{{\mathbf P}}
\newcommand{\bfW}{{\mathbf W}}

\newcommand{\bmeps}{\mbox{\boldmath $\epsilon$}}
\newcommand{\rmd}{{\mathrm d}}
\newcommand{\tilk}{{\tilde k}}
\newcommand{\tilK}{{\tilde K}}

\DeclarePairedDelimiterX{\inp}[2]{\langle}{\rangle}{#1, #2}
\DeclareMathOperator*{\diag}{diag}

\DeclareMathOperator*{\argmin}{argmin}
\DeclareMathOperator*{\argmax}{argmax}

\newcommand{\pluseq}{\mathrel{+}=}

\crefname{equation}{}{}
\crefname{figure}{Fig.}{Figs.}
\crefname{table}{Table}{Tables}
\crefname{algorithm}{Algorithm}{Algorithms}
\crefname{algocf}{Algorithm}{Algorithms}
\Crefname{algocf}{Algorithm}{Algorithms}
\crefname{section}{Section}{Sections}
\crefname{subsection}{Section}{Sections}

\theoremstyle{definition}
\newtheorem{theorem}{Theorem}[section]
\crefname{theorem}{Theorem}{Theorems}

\crefname{lemma}{Lemma}{Lemmas}

\crefname{corollary}{Corollary}{Corollaries}
\newtheorem{proposition}[theorem]{Proposition}
\crefname{proposition}{Proposition}{Propositions}

\crefname{definition}{Definition}{Definitions}
\newtheorem{remark}[theorem]{Remark}
\crefname{remark}{Remark}{Remarks}

\crefname{assumption}{Assumption}{Assumptions}

\crefname{example}{Example}{Examples}

\makeatletter
\newenvironment{refproof}[1][\proofname]{\par
  \pushQED{\qed}%
  \normalfont \topsep6\p@\@plus6\p@\relax
  \trivlist
  \item[\hskip\labelsep
         \bfseries
    Proof of {#1}]\ignorespaces
}{%
  \popQED\endtrivlist\@endpefalse
}
\makeatother

\title{Reservoir Predictive Path Integral Control for\\ Unknown Nonlinear Dynamics}

\blind{}{
    \author{Daisuke Inoue, Tadayoshi Matsumori, Gouhei Tanaka,~\IEEEmembership{Member,~IEEE}, and Yuji Ito,~\IEEEmembership{Senior Member,~IEEE}
        \thanks{Corresponding authors: Daisuke Inoue, Yuji Ito}
        \thanks{This study is partly supported by JSPS KAKENHI Grant Numbers JP23K28154, JP25H00451, JST Moonshot R\&D Grant Number JPMJMS2021, and JST CREST Grant Number JPMJCR24R2.}
        \thanks{D.~Inoue, T.~Matsumori, and Y.~Ito are with Toyota Central R\&D Labs., Inc., Nagakute, Aichi 480-1192, Japan}
        \thanks{G.~Tanaka is with Nagoya Institute of Technology, Nagoya, Aichi, 466-8555, Japan}
    }   
}

\preprint{}{
    \markboth{IEEE Transactions on Neural Networks and Learning Systems}{}
}

\begin{document}

\maketitle

\preprint{
    \IEEEpubid{\textbf{This work has been submitted to the IEEE for possible publication. Copyright may be transferred without notice, after which this version may no longer be accessible.}}
}{
    \IEEEpubid{0000--0000/00\$00.00~\copyright~2021 IEEE}
}


\begin{abstract}
    Neural networks have found extensive application in data-driven control of nonlinear dynamical systems, yet fast online identification and control of unknown dynamics remain central challenges.
    To meet these challenges, this paper integrates echo-state networks (ESNs)---reservoir computing models implemented with recurrent neural networks---and model predictive path integral (MPPI) control---sampling-based variants of model predictive control.
    The proposed reservoir predictive path integral (RPPI) enables fast learning of nonlinear dynamics with ESNs and exploits the learned nonlinearities directly in MPPI control computation without linearization approximations.
    This framework is further extended to uncertainty-aware RPPI (URPPI), which achieves robust stochastic control by treating ESN output weights as random variables and minimizing an expected cost over their distribution to account for identification errors.
    Experiments on controlling a Duffing oscillator and a four-tank system demonstrate that URPPI improves control performance, reducing control costs by up to 60\% compared to traditional quadratic programming-based model predictive control methods.
\end{abstract}


\section{Introduction}\label{sec:intro}

\preprint{
    \IEEEpubidadjcol
}{
    \IEEEpubidadjcol
}

\IEEEPARstart{A}{chieving} high-precision, real-time control of systems with unknown dynamics is challenging in many industrial applications.
Standard approaches first estimate the system model through identification, then design model-based control~\cite{Ljung1999System,Lewis2012Optimal}.
When the target system is linear, this approach can achieve high control performance based on mature identification and control theory.
However, real-world systems exhibit nonlinearities, making linear theory inadequate and degrading performance~\cite{nijmeijer1990nonlinear}.
To address this, machine learning-based modeling integrated with model-based control has gained much attention~\cite{Pan2008Nonlinear,Chen2018Optimala,Yan2012Model,Bonassi2021Nonlinear,Terzi2021Learning}.
As such implementations, neural networks (NNs)~\cite{Pan2008Nonlinear,Chen2018Optimala,Yan2012Model,Bonassi2021Nonlinear,Terzi2021Learning} and Gaussian process regression (GPR)~\cite{Cao2017Gaussiana,Polcz2023Efficient} with nonlinear model predictive control (MPC) show promise for nonlinear systems.
NNs deliver accurate identification of nonlinear dynamics and GPR enables uncertainty-aware control.
This study focuses on integrating data-driven models with MPC to enable online control of unknown nonlinear systems.

While combining machine learning models with nonlinear MPC has shown great promise for controlling unknown nonlinear systems, such approaches often face computational challenges in scenarios demanding fast online learning.
In practice, NN-based approaches incur excessive computational overhead in learning model parameters and solving optimization in nonlinear MPC~\cite{Salzmann2023RealTime}.
GPR-based methods typically suffer from computational complexity during identification that scales cubically with the number of training samples~\cite{williams2006gaussian,Park2020Gaussian}.
Although explicit MPC mitigates these challenges by pre-computing optimal input mappings, offline strategies are infeasible for unknown systems requiring online model updates~\cite{Alessio2009Survey,Kvasnica2013Complexity,Chen2018Approximating}.
Consequently, the development of methods that achieve both fast online identification and control remains a fundamental challenge in nonlinear control~\cite{Forssell1999Closedloop,Qin2006overview,Kvasnica2013Complexity}.

Reservoir computing, a framework for efficient neural computation, offers fast training and strong temporal modeling capabilities; among its variants, echo state networks (ESNs) are particularly notable for their simplicity and rapid adaptation~\cite{Jaeger2002Adaptive,Jaeger2007Optimization,Salmen2005Echo}.
Unlike end-to-end trained recurrent NNs, ESNs offer computational efficiency in online training, since the recurrent (reservoir) weights are fixed and only output weights need to be learned~\cite{Sun2024Systematic}.
Using recursive least squares (RLS), the output weights are updated with a constant computational cost per step, independent of the number of training samples.
Among nonlinear MPC approaches combined with ESN, quadratic programming MPC (QPMPC) has gained attention because of its computational efficiency~\cite{Pan2012Model,Jordanou2022Echo,Schwedersky2022Adaptive,Schwedersky2022Echo}.
This method linearizes a plant model at each control step, casting the optimization problem as a quadratic programming (QP) problem to enable rapid computation of the optimal control inputs.
This approach facilitates online computation of control inputs for nonlinear systems in a computationally efficient manner.

\preprint{
    \IEEEpubidadjcol
}{
    \IEEEpubidadjcol
}

However, ESN-based QPMPC approaches exhibit two limitations.
First, linearizing ESN models and restricting a cost function to a quadratic form both limit the method's applicability.
Because the ESN is linearized at every control step, its nonlinear expressive power cannot be fully exploited, leading to large prediction errors when the plant exhibits strong nonlinearity.
Furthermore, constraining the cost function to a quadratic form prevents the inclusion of higher-order terms required by certain control objectives.
Second, earlier studies~\cite{Pan2012Model,Jordanou2022Echo,Schwedersky2022Adaptive,Schwedersky2022Echo} typically rely on point estimates of model parameters, neglecting the uncertainty inherent in online identification.
In online scenarios where data is limited, the estimated model may deviate from the true dynamics.
Controllers based solely on such nominal models often degrade control performance when the model is inaccurate.
Thus, developing control methods that address these limitations remains a pressing challenge.

\begin{figure*}[t]
    \centering
    \includegraphics[width=\linewidth]{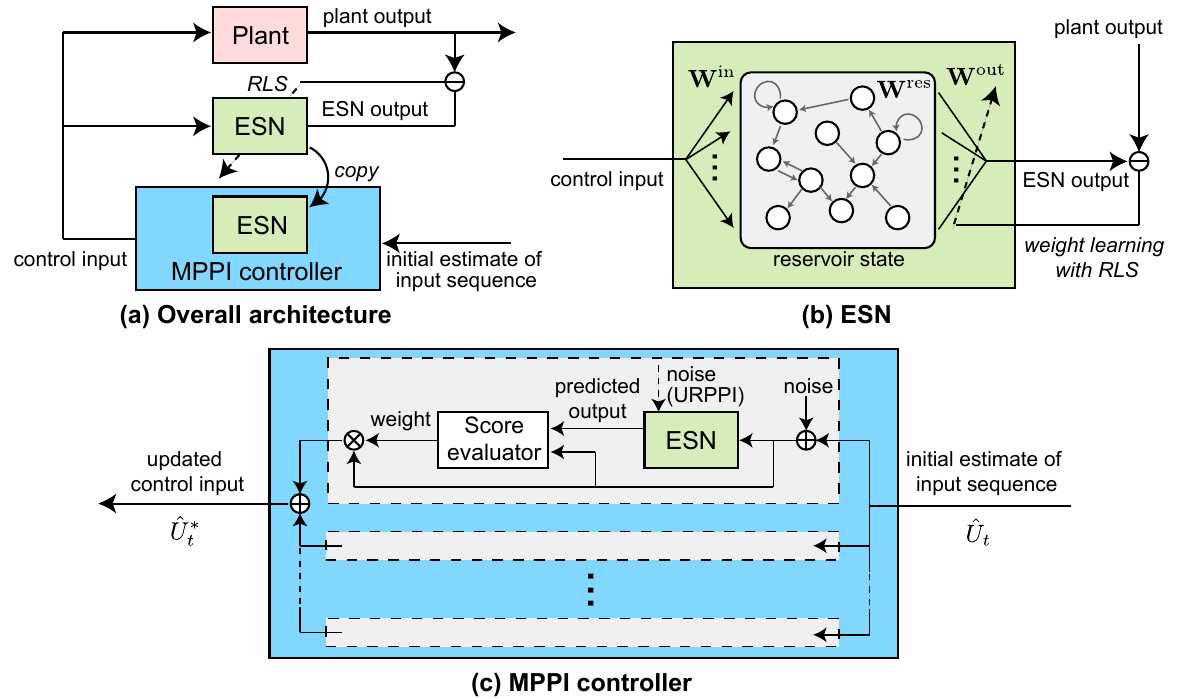}
    \caption{Schematic diagram of the proposed RPPI that integrates ESN and MPPI.
        (a) Overall architecture of RPPI; (b) ESN model for online identification of nonlinear dynamics, in which only the output weight matrix is trained with RLS; (c) MPPI controller for computing control inputs based on the learned ESN models. In URPPI, the output is predicted using the perturbed output weight matrices so that the controller can compute control inputs that minimize the expected cost under model uncertainty.
    }
    \label{fig:schematics}
\end{figure*}

To address the above challenges, we propose a control method called \emph{reservoir predictive path integral (RPPI)}, which combines an ESN with model predictive path integral (MPPI) control~\cite{Williams2018InformationTheoretic}.
MPPI is a sampling-based control method that generates numerous input trajectories and computes their weighted average to determine the control input.
This approach directly treats nonlinear dynamics and arbitrary cost functions without requiring explicit optimization solvers~\cite{Williams2015Model}.
Real-time computation has also been shown to be possible through parallel sampling using GPUs~\cite{Kazim2024Recent}.
Furthermore, we propose \emph{uncertainty-aware RPPI (URPPI)}, which incorporates model uncertainty into MPPI input weighting.
The model uncertainty is quantified using the covariance matrix associated with the RLS estimation of the ESN output weights.
This mechanism allows the controller to minimize the expected cost over the distribution of possible system dynamics, leading to control inputs that are robust to identification errors.
The schematic diagram of the proposed method is shown in \cref{fig:schematics}.
The main contributions of this research are summarized as follows:
\begin{itemize}
    \item \textbf{Online MPC for unknown nonlinear dynamics}: By leveraging ESN-based fast online learning, we propose an RPPI control framework that enables online MPC for strongly nonlinear unknown systems and arbitrary cost functions \emph{without} linearization.

    \item \textbf{Uncertainty-aware stochastic control}: We extend RPPI to a URPPI by injecting the ESN output-weight covariance into the MPPI weighting mechanism, which enables stochastic optimal control that approximately minimizes the expected cost under model uncertainty.

    \item \textbf{Experimental validation}: We demonstrate the performance of the proposed approach on the Duffing oscillator and four-tank system benchmarks, where URPPI achieves up to 60\% lower control cost than ESN-based QPMPC.
\end{itemize}

In the following, \cref{sec:related} reviews related work on ESN-based MPC and machine learning-based control.
\cref{sec:problem} presents the problem setting of this research.
Next, \cref{sec:proposed-method} introduces the ESN model and its learning algorithm, RLS, and then proposes RPPI and URPPI as methods that integrate ESN with MPPI.
In \cref{sec:numerical_experiments}, we conduct numerical simulations to evaluate the performance of the proposed methods.
Finally, conclusions and directions for future work are discussed in \cref{sec:conclusion}.

\textbf{Notation.}
For a matrix $A\in\bbR^{m\times n}$, the vectorization $\mathrm{vec}(A)$ is defined as $\mathrm{vec}(A)\coloneqq[a_{11},\ldots,a_{m1},\ldots,a_{1n},\ldots,a_{mn}]^\top\in\bbR^{mn}$, where $a_{ij}$ denotes the $(i,j)$-th entry of $A$.
For matrices $A\in\bbR^{m\times n}$ and $B\in\bbR^{p\times q}$, the Kronecker product $A\otimes B\in\bbR^{mp\times nq}$ is defined as
\begin{align}\nonumber
    A\otimes B \coloneqq \begin{bmatrix}
                             a_{11} B & \cdots & a_{1n} B \\
                             \vdots   & \ddots & \vdots   \\
                             a_{m1} B & \cdots & a_{mn} B
                         \end{bmatrix}.
\end{align}
For a random variable $v$ following a distribution $p$, the expectation of a function $f$ is denoted by $\bbE_{p}[f(v)]$.
The symbol $\calN(\mu, \Sigma)$ denotes the multivariate normal distribution with mean $\mu$ and covariance matrix $\Sigma$, and its density function is written as $\calN(\cdot | \mu, \Sigma)$.

\section{Related Work}\label{sec:related}

Several ESN-based MPC approaches have been proposed, each with distinct characteristics and limitations.
Pan and Wang~\cite{Pan2012Model} introduced ESNs as surrogate models for nonlinear MPC by training the ESN offline and applying a first-order Taylor expansion to convexify the optimization into QP; because the model is fixed and locally linearized, the controller struggles with strong nonlinear dynamics and cannot adapt to new data.
Jordanou \emph{et al.}~\cite{Jordanou2022Echo} preserved part of the ESN's nonlinearity by separating forced and free responses, yet they still linearize and solve a QP at every step.
Schwedersky \emph{et al.}~\cite{Schwedersky2022Adaptive} added RLS updates so that the ESN adapts online, but their controller linearizes the ESN and solves a QP each iteration.
In contrast, the proposed \emph{RPPI/URPPI} framework eliminates \emph{all} linearization approximations through sampling-based path-integral control.

Learning-based control methods face a trade-off between model expressiveness and computational efficiency.
NN models, including recurrent NN (RNN), gated recurrent units (GRU), long short-term memory (LSTM), transformer model, input convex NN (ICNN), and graph NN (GNN), offer high expressiveness but require computationally expensive gradient computations for online learning~\cite{Pan2008Nonlinear,Chen2018Optimala,Yan2012Model,Bonassi2021Nonlinear,Terzi2021Learning,Sanchez-Gonzalez2018Graph,Park2023Simultaneous,Wang2024Fast}.
GPR-based approaches provide uncertainty quantification but typically suffer from computational complexity that scales with the accumulated training data points $N_{\text{data}}$ (e.g., $O(N_{\text{data}}^3)$ for standard GPR), which limits their applicability for long-term online learning~\cite{Cao2017Gaussiana,Polcz2023Efficient}.
In contrast, the ESN architecture balances nonlinear modeling capability with computational efficiency.
By adapting only the linear readout weights via RLS, ESN achieves a constant computational cost of $O(\hat N^2)$ per time step, independent of the number of processed data points, where $\hat N$ is the reservoir state dimension.

Reinforcement learning is another powerful approach for controlling systems with unknown dynamics~\cite{MoerlandModelbased2023,Brunke2022Safe}.
Probabilistic ensembles with trajectory sampling (PETS) utilizes NN ensembles to minimize expected costs~\cite{Chua2018Deep}, and imitation learning methods approximate stochastic control laws to reduce computational limits~\cite{Pozzi2025Imitation}.
Although these methods demonstrate high performance in episodic settings with abundant offline training data or expert demonstrations, this paper focuses on a different scenario: strictly online control starting from scratch.
The proposed URPPI framework leverages ESN's analytical posterior updates for online stochastic optimal control, enabling online uncertainty quantification and robust adaptation within a single execution.

Our control strategy is based on MPPI, a sampling-based framework that enables efficient computation of control inputs for complex and non-differentiable dynamics~\cite{Williams2018InformationTheoretic}.
We extend this foundation to incorporate model uncertainty.
While other uncertainty-aware approaches exist, such as tube-based or risk-aware MPPI~\cite{williams2018robust, Yin2023RiskAware}, our approach utilizes a computationally efficient mechanism.
Specifically, we use the covariance matrix of the ESN's output weights, obtained as a natural byproduct of the RLS learning process, as a measure of parametric uncertainty.
As a result, the controller can consider the distribution of possible models and compute inputs that are robust on average.

\section{Problem Formulation}\label{sec:problem}

In this paper, we consider a control system with input dimension $M\in\bbN$, state dimension $N\in\bbN$, and output dimension $L\in\bbN$, described as follows:
\begin{align}
    \bfz_{t+1} & =\bfF(\bfz_{t}, \bfu_{t}),\label{eq:plant} \\
    \bfy_{t}   & =\bfG(\bfz_{t})\label{eq:plant-output},
\end{align}
where $\bfz_{t}\in\bbR^N$ is the state vector, $\bfu_{t}\in\bbR^M$ is the input vector, and $\bfy_{t}\in\bbR^L$ is the output vector.
The symbol $\bfF\colon\bbR^N\times\bbR^M\to\bbR^N$ and $\bfG\colon\bbR^N\to\bbR^L$ are the state transition function and output function, respectively.

For such a control system, we consider the problem of finding a control input sequence $U_t\in\bbR^{M\times H}$ that
minimizes the following cost function $J_\text{ctrl}$ at each time $t$:
\begin{align}
     & \begin{aligned}
           J_\text{ctrl}(U_t; \bfz_t) & \coloneqq C\left(\bfy_t, \bfy_{t+1}, \ldots \bfy_{t+H}\right)                                           \\
                                      & \quad +\sum_{\tau=t}^{t+H-1} \frac{1}{2}(\bfu_\tau-\bfu^\text{ref})^\top R (\bfu_\tau-\bfu^\text{ref}),
       \end{aligned}\label{eq:cost-function-def}
\end{align}
where $H\in\bbN$ is the prediction horizon and the output sequence $\bfy_{t}, \ldots, \bfy_{t+H}$ is uniquely determined by the initial state $\bfz_t$ and the control input sequence $U_t$ via the system dynamics \cref{eq:plant,eq:plant-output}.
The input sequence $U_t$ consists of control input vectors $\bfu_\tau$ for $\tau=t,\ldots,t+H-1$.
The matrix $R\in\bbR^{M\times M}$ is a positive definite weight matrix, and $\bfu^\text{ref}\in\bbR^M$ denotes the reference control input.
The function $C$ is a cost function related to the time series of the output vectors, expressed using running cost functions $\ell\colon\bbR^L\to\bbR$ and terminal cost function $\phi\colon\bbR^L\to\bbR$ as follows:
\begin{align}
    C(\bfy_t, \bfy_{t+1}, \ldots \bfy_{t+H}) \coloneqq \sum_{\tau=t}^{t+H-1} \ell(\bfy_\tau)  + \phi(\bfy_{t+H}).
\end{align}

In this paper, we assume that only the input-output pair $(\bfu_t, \bfy_t)$ is observable at each time step $t$, while the state transition function $\bfF$ and output function $\bfG$ in \cref{eq:plant,eq:plant-output} are unknown.
To account for the lack of knowledge, we formulate the control problem as a stochastic MPC problem.
We address the problem of finding a control input sequence $U_t$ that minimizes the expected cost with respect to the parameter $\bfw_t$ representing the model uncertainty:
\begin{align}
     & \underset{U_t\in \calU}{\text{minimize}}\ \bbE_{\mathcal{P}_t} [J_\text{ctrl}(U_t)],     \label{eq:stochastic-cost-function} \\
     & \text{subject to}\ \bfx_{t+1} = \hat{\bfF}(\bfx_t, \bfu_t; \bfw_t), \label{eq:stochastic-dynamics-state}                     \\
     & \phantom{\text{subject to}\ }\hat{\bfy}_{t\phantom{+1}} = \hat{\bfG}(\bfx_t; \bfw_t). \label{eq:stochastic-dynamics-output}
\end{align}
Here, the system dynamics are approximated by a surrogate model $\hat\bfF$ and $\hat\bfG$ parameterized by $\bfw_t$.
The symbol $\mathcal{P}_t$ denotes the probability distribution of the parameter $\bfw_t$.
The variable $\bfx_t$ denotes the state of the surrogate model, and $\hat{\bfy}_t\in\bbR^L$ denotes its output.
The cost function $J_\text{ctrl}(U_t)$ in \cref{eq:stochastic-cost-function} is evaluated along the trajectory generated by this surrogate model.
The symbol $\calU\subseteq\bbR^{M\times H}$ is the set of admissible control inputs.
As reviewed in Ref.~\cite{Mesbah2018Stochastic}, this stochastic formulation enables the controller to account for model uncertainty, leading to a robust control strategy that reduces the risk of performance degradation due to identification errors.
The goal of this paper is to solve this stochastic MPC problem online while simultaneously identifying the system dynamics from observed data.

Solving \cref{eq:stochastic-cost-function} presents two major challenges.
First, obtaining the surrogate model of the system dynamics, including the uncertainty distribution $\mathcal{P}_t$, from limited data online is difficult for general nonlinear models.
Second, minimizing the cost function is computationally intractable, as it results in a non-convex optimization problem when the model is nonlinear and the cost function includes high-order terms.
Our aim is to overcome these challenges.

\section{Proposed Method}\label{sec:proposed-method}

\begin{algorithm}[t]
    \caption{RPPI/URPPI}\label{alg:RPPI}
    \SetKwInOut{Given}{Given}
    \Given{Initial reservoir state $\bfx_0$, initial estimate of input sequence $\hat U_{-1}$, terminal time $T$,
    initial output weight $\bfW_{-1}^{\text{out}}$ and precision matrix $\bfP_{-1}$.\\
    }
    $t\leftarrow 0$.\\
    \While{$t<T$}{
    Update the ESN output weight $\bfW^{\text{out}}_t$ and the precision matrix $\bfP_t$ using RLS (\cref{alg:RLS}) with the observed plant output $\bfy_t$, reservoir state $\bfx_t$, previous output weight $\bfW_{t-1}^{\text{out}}$, and precision matrix $\bfP_{t-1}$.\\
    Calculate the control input sequence $\hat U^*_t$ with MPPI/UMPPI (\cref{alg:UMPPI}), using previous sequence $\hat U_t=\hat U^*_{t-1}$, reservoir state $\bfx_t$, output weight $\bfW^{\text{out}}_t$ and precision matrix $\bfP_t$.\\
    Apply the first component $\hat \bfu^*_t$ of $\hat U^*_t$ to both the ESN and plant to update each state $\bfx_{t+1}$ and $\bfz_{t+1}$.\\
    $t\leftarrow t+1$.
    }
\end{algorithm}

This section presents a framework that integrates the ESN with MPPI.
To overcome the difficulty of online probabilistic modeling, we employ an ESN trained with RLS.
Distinct from standard deterministic approaches, we utilize the precision matrix estimated by RLS to quantify the posterior variance of the model parameters, providing a lightweight and analytical expression for model uncertainty.
To address the intractability of non-convex optimization, we adopt the MPPI framework.
This method approximates the optimal control problem as a free-energy minimization that can be solved efficiently via importance sampling without linearization approximations.
The integration of these components results in the URPPI algorithm, providing a unified framework for online stochastic optimal control of unknown nonlinear systems.

The proposed methods, RPPI and URPPI, operate through the following three-step cycle at each control iteration:
\begin{enumerate}
    \item Update the ESN output weight matrix $\bfW^{\text{out}}_t$ and its precision matrix $\bfP_t$ using RLS.
    \item Compute the control input sequence $\hat U^*_t$ using MPPI or UMPPI, where UMPPI extends MPPI by injecting ESN output weight uncertainty into the sampling process.
    \item Update the reservoir state $\bfx_t$ and control system state $\bfz_t$ by applying input $\hat\bfu^*_t$, the first component of $\hat U^*_t$.
\end{enumerate}
An overview is shown in \cref{alg:RPPI}.
The remainder of this section details the components of the proposed framework.
\Cref{sec:ESN} describes the ESN architecture and online learning using RLS.
\Cref{sec:MPPI} explains the MPPI-based control using the ESN model.
\Cref{sec:complexity} evaluates the computational complexity of the overall procedure.

\subsection{Echo State Network}\label{sec:ESN}

We describe the ESN model and its learning method.
The ESN is a lightweight model capable of approximating complex nonlinear dynamics.
We employ the ESN as a probabilistic model to quantify the uncertainty of the learned dynamics, and employ RLS~\cite{farhang2013adaptive,haykin2002adaptive} for online learning of the model parameters.

\subsubsection{Model and Learning Algorithm}

As a surrogate model in \cref{eq:stochastic-dynamics-state,eq:stochastic-dynamics-output}, we employ the ESN with leaky terms, which incorporates temporal dynamics through leak rates of the target control system~\cite{Jaeger2007Optimization}:
\begin{align}
     & \begin{aligned}
           \bfx_{t+1} & = \hat\bfF(\bfx_t, \bfu_t)                                                                                  \\
                      & \coloneqq (1-\alpha) \bfx_t + \alpha\, f \left(\bfW^{\text {res}} \bfx_t + \bfW^{\text {in }}\bfu_t\right),
       \end{aligned} \label{eq:esn} \\
     & \hat \bfy_t =\bfW^{\text{out} \top}_t \bfx_t.\label{eq:esn_output}
\end{align}
Here, $\bfx_t\in\bbR^{\hat N}$ is the reservoir state vector, where $\hat N\in\bbN$, and $\hat \bfy_t\in\bbR^{L}$ is the output vector.
Additionally, $\bfW^{\text{res}}\in\bbR^{\hat N\times \hat N}$ and $\bfW^{\text{in}}\in\bbR^{\hat N\times M}$ are the state and input weight matrices, respectively, and
$\bfW^{\text{out}}_t\in\bbR^{\hat N\times L}$ is the output weight matrix.
Furthermore, $f\colon\bbR^{\hat N}\to\bbR^{\hat N}$ is an element-wise nonlinear activation function, and $\alpha\in (0,1]$ is a parameter called the leak rate.

The ESN is a lightweight model that enables fast learning of dynamical systems with nonlinear characteristics.
This computational efficiency arises because, unlike conventional NNs that train all weight matrices, only the output weight matrix $\bfW^{\text{out}}_t$ needs to be trained.
The state and input weight matrices $\bfW^{\text {res}}$ and $\bfW^{\text{in}}$ remain fixed as randomly initialized matrices.
The reservoir state dimension $\hat N$ is set greater than the input or output dimensions.
The matrix $\bfW^{\text {res}}$ is a sparse random matrix with many components equal to zero, designed empirically so that the maximum absolute value of eigenvalues (spectral radius) is 1 or less.
The matrix $\bfW^{\text{in}}$ is randomly generated from a uniform distribution.
The reservoir state dimension $\hat N$, sparsity and spectral radius of $\bfW^{\text{res}}$, the shape of the activation function $f$, the leak rate $\alpha$, and the range of the input weight matrix $\bfW^{\text{in}}$ are hyperparameters that must be set prior to learning.

In online settings where data arrives sequentially, the ESN enables particularly rapid learning through RLS.
This algorithm minimizes the following weighted least squares error:
\begin{align}\label{eq:cost-function-RLS}
    J_\text{id}(\bfW^{\text{out}}) = \sum_{\tau=0}^{t} \gamma^{t-\tau}\|\bfy_\tau - \bfW^{\text{out} \top} \bfx_\tau\|_2^2,
\end{align}
where $\gamma\in(0,1]$ is the discount factor.
The following proposition summarizes the recursive update rule.

\begin{proposition}[Ref.~\cite{farhang2013adaptive}]\label{prop:RLS}
    Assume that for every time $t\ge \hat N-1$, the matrix $\bfA_t\coloneqq \sum_{\tau=0}^t \gamma^{t-\tau} \bfx_\tau \bfx_\tau^\top$ is invertible.
    Then, the minimizer $\bfW^{\text{out}*}_{t}$ at time $t\ge \hat N$ can be expressed using
    the minimizer $\bfW^{\text{out}*}_{t-1}$ at time $t-1$ as follows:
    \begin{align}
        \bfW^{\text{out}*}_t & =\bfW^{\text{out}*}_{t-1}+\frac{\bfP_{t-1} \bfx_t}{\gamma+\bfx_t^{\top} \bfP_{t-1} \bfx_t} \left(\bfy_t-\bfW^{\text{out}* \top}_{t-1} \bfx_t\right)^\top, \label{eq:RLS_W} \\
        \bfP_t               & =\frac{1}{\gamma}\left(\bfP_{t-1}-\frac{\bfP_{t-1} \bfx_t \bfx^{\top}_t \bfP_{t-1}}{\gamma+\bfx_t^{\top} \bfP_{t-1} \bfx_t}\right), \label{eq:RLS_P}
    \end{align}
    where $\bfP_t = \bfA_t^{-1} \in\bbR^{\hat N\times \hat N}$ is called the precision matrix of the variable $\bfx_t$.
\end{proposition}

\begin{algorithm}[t]
    \caption{RLS~\cite{farhang2013adaptive}}\label{alg:RLS}
    \SetKwInOut{Given}{Given}
    \Given{Discount factor $\gamma$,
        previous output weight $\bfW^{\text{out}}_{t-1}$, precision matrix $\bfP_{t-1}$,
        observed output $\bfy_t$, and current state $\bfx_t$.
    }
    $\bfW^{\text{out}}_t \leftarrow \bfW^{\text{out}}_{t-1}+\frac{\bfP_{t-1} \bfx_t}{\gamma+\bfx_t^{\top} \bfP_{t-1} \bfx_t} \left(\bfy_t-\bfW^{\text{out} \top}_{t-1} \bfx_t\right)^\top$.\\
    $\bfP_t \leftarrow \frac{1}{\gamma}\left(\bfP_{t-1}-\frac{\bfP_{t-1} \bfx_t \bfx^{\top}_t \bfP_{t-1}}{\gamma+\bfx_t^{\top} \bfP_{t-1} \bfx_t}\right)$.\\
    \Return{$\bfW^{\text{out}}_t$ and $\bfP_t$}
\end{algorithm}

The sequential learning method that repeatedly applies \cref{eq:RLS_W,eq:RLS_P} with arbitrarily initialized parameters $\bfW_{-1}^{\text{out}}$ and $\bfP_{-1}$ is called RLS~\cite{farhang2013adaptive,haykin2002adaptive}.
The algorithm is shown in \cref{alg:RLS}.
In this algorithm, the computational cost required at each step is $O(\hat N^2+L\hat N)$ for matrix multiplication operations, which is independent of the number of past data points.
This constant computational cost makes RLS suitable for online learning scenarios where data arrives sequentially.

\subsubsection{Probabilistic Interpretation}

We treat the ESN as a probabilistic model to quantify the uncertainty of the model during learning.
Given that $\bfW^{\text{out}}_t$ is the only learnable parameter in the ESN, we examine the probability distribution of $\bfW^{\text{out}}_t$ and provide a probabilistic interpretation of \cref{eq:esn,eq:esn_output}.

In RLS, $\bfW^{\text{out}}_t$ is updated sequentially with each newly observed data.
In the early stages of learning, the estimation error is large due to the limited amount of observed data.
As the amount of data becomes sufficiently large, the RLS estimate converges to the ordinary least squares estimator.
Therefore, we define $\bar \bfW^{\text{out}}$ as the limit of $\bfW^{\text{out}}_t$ as $t\to\infty$ and define the learning error as follows:
\begin{align}\label{eq:output_true}
    \bmeps_t \coloneqq \bfy_t-\bar \bfW^{\text{out} \top}\bfx_t.
\end{align}
We assume that this error $\bmeps_t\in\bbR^L$ behaves as stochastic noise with mean $0$ and covariance matrix $\tilde\Sigma=\diag(\sigma_1^2,\ldots,\sigma_L^2)$ independently at each time.
Although this assumption acts as an approximation for the complex error dynamics in practice, it allows us to derive the following probabilistic properties of $\bfW^{\text{out}}_t$.

\begin{proposition}\label{prop:RLS_covariance}
    If $\gamma=1$,
    \begin{align}
        \bbE[\mathrm{vec}(\bfW^{\text{out}*}_t - \bar \bfW^{\text{out}}) \mathrm{vec}(\bfW^{\text{out}*}_t - \bar \bfW^{\text{out}})^{\top}] = \tilde\Sigma\otimes\bfP_t.
    \end{align}
\end{proposition}
This property allows us to quantify the model uncertainty $\mathcal{P}_t$ introduced in \cref{eq:stochastic-cost-function}.
By stochastically sampling the output weight matrix using $\bfP_t$, we can approximately evaluate the expected value in \cref{eq:stochastic-cost-function}.
Furthermore, since the ESN represents uncertainty solely through the output weight matrix, this sampling strategy provides computational advantages over methods that propagate uncertainty through the entire state vector (see Remark \ref{rem:complexity}).
Note that Proposition \ref{prop:RLS_covariance} is obtained by extending the analysis in Ref.~\cite{farhang2013adaptive} to the case of ESNs, where the estimated parameters take a matrix form; a detailed proof is provided in the Appendix.

\subsection{Uncertainty-aware Model Predictive Path Integral Control}\label{sec:MPPI}

We propose control methods combining the ESN with MPPI or its uncertainty-aware variant UMPPI.
In MPPI~\cite{Williams2018InformationTheoretic}, multiple noisy input sequences are sampled based on a reference input to simulate future trajectories of the control system, and an output cost is computed for each sequence.
The optimal control input is approximately computed as a weighted average of many samples, where samples with lower costs receive higher weights.
In UMPPI, in addition to sampling input sequences, output weight matrices of the ESN with noise added are sampled.
This provides control input perturbations according to the identification accuracy, allowing the controller to account for model uncertainty and achieve robust performance.
Note that in \cref{alg:UMPPI} described below, when the sample size $\tilK=1$ and the noise covariance $\tilde\Sigma=0$, UMPPI reduces to the standard MPPI algorithm.
Therefore, we focus on the UMPPI in the following description.

\begin{remark}
    The following derivation of UMPPI extends the framework of Ref.~\cite{Williams2018InformationTheoretic} by introducing perturbations of the output-weight matrix.
    This extension preserves the mathematical consistency of the MPPI formulation and constitutes a theoretical contribution of the present work.
    Specifically, Proposition~\ref{prop:free_energy_inequality} parallels Eq.~(13) in Ref.~\cite{Williams2018InformationTheoretic}; Proposition~\ref{prop:optimal_distribution} is established by the same argument as in Section~III-A of that reference; and Proposition~\ref{prop:importance_sampled_optimal_input} corresponds to Eq.~(27) in the same reference.
\end{remark}

We outline how UMPPI computes the optimal control solution through a sequence of sampling-based approximations.
The derivation proceeds in the following three steps:
\begin{enumerate}
    \item \textbf{Stochastic reformulation.}
          The stochastic optimal control problem in \cref{eq:stochastic-cost-function} is approximated by introducing noise into the control inputs and model parameters, as expressed in~\cref{eq:optimal_input_perturbed}.

    \item \textbf{Free-energy relaxation.}
          Introducing a free-energy functional converts this problem into the minimization of a Kullback–Leibler divergence between two probability distributions, as shown in \cref{eq:free_energy_optimal_input}.

    \item \textbf{Sampling-based optimization.}
          The resulting optimal input sequence is then approximated by Monte Carlo integration with importance sampling, as implemented in~\cref{alg:UMPPI}.
\end{enumerate}
The details of each step are described in \cref{sec:mppi-objective,sec:mppi-free-energy,sec:mppi-optimal-input}.

\subsubsection{Stochastic reformulation}\label{sec:mppi-objective}

This subsection reformulates the original MPC problem in \cref{eq:stochastic-cost-function} as an optimization problem over perturbed control inputs and model parameters in \cref{eq:optimal_input_perturbed}.
Like conventional MPPI, UMPPI considers perturbed inputs $\bfv_t\sim\calN(\bfu_t, \Sigma)$ where Gaussian noise is applied to some initial inputs $\bfu_t$:
\begin{align}\label{eq:esn_fluctuated}
    \hat\bfx_{\tau+1}=\hat \bfF(\hat\bfx_{\tau}, \bfv_{\tau}),\ \hat\bfx_t=\bfx_t,
\end{align}
where $\hat \bfF$ is defined in \cref{eq:esn}.
Furthermore, $\Sigma\in\bbR^{M\times M}$ represents the covariance matrix of the applied noise, and its value is determined based on the input weight matrix $R$ of the cost function \eqref{eq:cost-function-def}:
\begin{align}
    \Sigma = \lambda R^{-1},
\end{align}
where $\lambda\in\bbR$ is a parameter called the inverse temperature, which is a design parameter.
Additionally, UMPPI considers the output weight matrices $\bfW_\tau$ with noise applied to the reference weight matrix $\bfW^{\text{out}}_\tau$ at each time:
\begin{align}\label{eq:esn_output_fluctuated}
    \hat\bfy_{\tau} = \bfW_\tau^\top \hat\bfx_\tau,
\end{align}
where $\mathrm{vec}(\bfW_\tau)\sim\calN(\mathrm{vec}(\bfW^{\text{out}}_{t}), \tilde\Sigma\otimes\bfP_t)$, which is justified by Proposition \ref{prop:RLS_covariance}.
Applying noise to inputs enables approximate computation of optimal inputs through the relationship with free energy, as explained in the following subsections.
Additionally, applying noise to the output weight matrix allows us to approximate the expectation $\bbE_{\mathcal{P}_t}[\cdot]$ in \eqref{eq:stochastic-cost-function}, which results in a control strategy robust to identification errors.

For the input sequence to be determined, $U_t=(\bfu_t,\ldots,\bfu_{t+H-1})\in\bbR^{M\times H}$, perturbed input sequence $V_t = (\bfv_t,\ldots,\bfv_{t+H-1})\in\bbR^{M\times H}$, and output weight matrix sequence $W_t = (\bfW_t,\ldots,\bfW_{t+H})\in\bbR^{\hat N\times L \times (H+1)}$, the joint probability density function $p_{U_t}$ of $V_t$ and $W_t$ is expressed as
\begin{align}
     & \begin{aligned}\label{eq:density-vw}
           p_{U_t}(V_t,W_t) & = p(V_t|U_t,\Sigma)p(W_t|\bfW_t^{\text{out}},\bfP_t,\tilde\Sigma),
       \end{aligned}                                        \\
     & p(V_t|U_t,\Sigma)                               = \prod_{\tau=t}^{t+H-1} \calN(\bfv_\tau|\bfu_\tau,\Sigma), \label{eq:prob_V}                 \\
     & \begin{aligned}\label{eq:prob_W}
            & p(W_t|\bfW_t^{\text{out}},\bfP_t,\tilde\Sigma)                                                                             \\
            & \quad = \prod_{\tau=t}^{t+H}\calN(\mathrm{vec}(\bfW_\tau)|\mathrm{vec}(\bfW^{\text{out}}_{t}), \tilde\Sigma\otimes\bfP_t).
       \end{aligned}
\end{align}
%
In UMPPI, the system dynamics are approximated by the ESN model, and at each time $t$, we seek the input sequence that minimizes the cost function in~\cref{eq:cost-function-def} in the sense of the expected value for $p_{U_t}$:
\begin{align}\label{eq:optimal_input_perturbed}
     & \underset{U_t\in \calU}{\text{minimize}}\ \bbE_{p_{U_t}} \left[\hat J_\text{ctrl}(V_t, W_t)\right],                                                \\
     & \begin{aligned}
           \hat J_\text{ctrl}(V_t, W_t) & \coloneqq  S\left(V_t,W_t ; \hat \bfx_t\right)                                                          \\
                                        & \quad +\sum_{\tau=t}^{t+H-1} \frac{1}{2}(\bfu_\tau-\bfu^\text{ref})^\top R (\bfu_\tau-\bfu^\text{ref}),
       \end{aligned}             \\
     & S\left(V_t,W_t ; \hat \bfx_t\right)\coloneqq C\left(\bfW^\top_t\hat\bfx_t, \bfW^\top_{t+1}\hat\bfF\left(\hat\bfx_t, \bfv_t\right),  \ldots\right).
\end{align}

\subsubsection{Free energy relaxation}\label{sec:mppi-free-energy}

This subsection introduces a variational relaxation of the stochastic optimization in \cref{eq:optimal_input_perturbed} using the concept of free energy, leading to an alternative minimization problem in \cref{eq:free_energy_optimal_input}.
We define the free energy $\calF$ as follows:
\begin{align}
    \begin{aligned}\label{eq:free_energy}
         & \calF\left(S, q, \hat\bfx_t, \lambda\right)                                                                                                      \\
         & \coloneq-\lambda \left\{ \log \left(\bbE_{q}\left[\exp \left(-\frac{1}{\lambda} S\left(V_t,W_t ; \hat\bfx_t\right)\right)\right]\right)\right\},
    \end{aligned}
\end{align}
where $q$ is a joint probability density function for
$V_t$ and $W_t$ which is called the reference density function:
\begin{align}
    q(V_t,W_t) & \coloneqq q(V_t)q(W_t),                                                                                                                            \\
    q(V_t)     & \coloneqq \prod_{\tau=t}^{t+H-1} \calN(\bfv_\tau|\bfu^\text{ref},\Sigma), \label{eq:base_prob_V}                                                   \\
    q(W_t)     & \coloneqq \prod_{\tau=t}^{t+H}\calN(\mathrm{vec}(\bfW_\tau)|\mathrm{vec}(\bfW^{\text{out}}_{t}), \tilde\Sigma\otimes\bfP_t).\label{eq:base_prob_W}
\end{align}
Here, $\bfu^\text{ref}$ is the reference input defined in \cref{eq:cost-function-def}.
The following two propositions show the relationship between \cref{eq:free_energy} and \cref{eq:optimal_input_perturbed}.

\begin{proposition}\label{prop:free_energy_inequality}
    For any $U_t\in\calU$, the free energy \eqref{eq:free_energy} achieves a lower bound of the cost function \eqref{eq:optimal_input_perturbed}:
    \begin{align}\label{eq:free_energy_inequality}
        \calF\left(S, q, \hat\bfx_t, \lambda\right) \le \bbE_{p_{U_t}} \left[\hat J_\text{ctrl}(V_t,W_t)\right].
    \end{align}
\end{proposition}

\begin{proposition}\label{prop:optimal_distribution}
    Suppose that $V_t$ and $W_t$ in \cref{eq:prob_V,eq:prob_W} follow the distributions given by
    \begin{align}
        p^*(V_t,W_t) & \coloneqq \frac{1}{\eta} \exp \left(-\frac{1}{\lambda} S\left(V_t,W_t ; \hat\bfx_t\right)\right) q(V_t, W_t), \label{eq:optimal_distribution} \\
        \eta         & \coloneqq \bbE_{q} \left[\exp \left(-\frac{1}{\lambda} S\left(V_t,W_t ; \hat\bfx_t\right)\right)\right].
    \end{align}
    Then, equality holds for \cref{eq:free_energy_inequality}:
    \begin{align}\label{eq:free_energy_optimal}
        \calF\left(S, q, \hat\bfx_t, \lambda\right) = \bbE_{p^*} \left[\hat J_\text{ctrl}(V_t,W_t)\right].
    \end{align}
\end{proposition}

From Propositions \ref{prop:free_energy_inequality} and \ref{prop:optimal_distribution}, designing the distributions of $V_t$ and $W_t$ to satisfy \eqref{eq:optimal_distribution} guarantees the optimality of the resulting control input sequence $U_t$.
Thus, we aim to find control inputs to achieve the distribution that minimizes the following Kullback-Leibler divergence:
\begin{align}
    \begin{aligned}
        U_t^{*}
         & =\underset{U_t \in \calU}{\argmin}\ \left\{\bbE_{p^*}\left[\log\frac{p^*(V_t,W_t)}{p_{U_t}(V_t,W_t)}\right]\right\}.
    \end{aligned}\label{eq:free_energy_optimal_input}
\end{align}

\subsubsection{Sampling-based optimization}\label{sec:mppi-optimal-input}

The following proposition provides a computational method for calculating the input \eqref{eq:free_energy_optimal_input}.
This is achieved using importance sampling based on some initial estimate of the input sequence $\hat U_t=(\hat\bfu_t,\ldots,\hat\bfu_{t+H-1})\in\bbR^{M\times H}$.
\begin{proposition}\label{prop:importance_sampled_optimal_input}
    Suppose that the admissible control input set is given by $\calU=\bbR^{M\times H}$.
    Then, given any input sequence $\hat U_t$,
    the control input satisfying \cref{eq:free_energy_optimal_input} is computed as
    \begin{align}\label{eq:importance_sampled_optimal_input}
         & \bfu_\tau^{*} =
        \frac{1}{\tilde\eta}\ \bbE_{p_{\hat U_t}}[w_t(V_t,W_t)\bfv_\tau],                                                                                                                                     \\
         & \begin{aligned}
               w_t(V_t,W_t) & \coloneqq \exp \left(-\frac{1}{\lambda}\Biggl(S\left(V_t,W_t ; \hat\bfx_t\right) \right.                                          \\
                            & \quad \left.\left.  +\sum_{\tau=t}^{t+H-1}(\hat\bfu_{\tau}-\bfu^{\text{ref}})^\top R (\bfv_\tau-\bfu^{\text{ref}})\right)\right),
           \end{aligned} \\
         & \tilde\eta \coloneqq \bbE_{p_{\hat U_t}}[w_t(V_t,W_t)],
    \end{align}
    where $p_{\hat U_t}$ represents the following distribution:
    \begin{align}
         & \begin{aligned}\label{eq:prob_U}
               p_{\hat U_t}(V_t,W_t) & \coloneqq p(V_t|\hat U_t,\Sigma)\, p(W_t|\bfW_t^{\text{out}},\bfP_t,\tilde\Sigma),
           \end{aligned}                \\
         & p(V_t|\hat U_t,\Sigma)                                             \coloneqq \prod_{\tau=t}^{t+H-1} \calN(\bfv_\tau|\hat\bfu_\tau,\Sigma).
    \end{align}
\end{proposition}

\begin{algorithm}[t]
    \caption{MPPI/UMPPI}\label{alg:UMPPI}
    \SetKwInOut{Given}{Given}
    \Given{Sample size of input/output perturbation $K,\tilK$,
        prediction horizon $H$,
        initial estimate of input sequence $\hat U_t = \hat U_{t-1}^*$,
        reservoir state $\bfx_t$, output weight $\bfW^{\text{out}}_t$, precision matrix $\bfP_t$,
        cost function $\ell$, $\phi$, inverse temperature $\lambda$, and noise covariances $\Sigma,\tilde\Sigma$.
    }
    $J_{k,\tilk}\leftarrow 0,\ k\in\{0,\ldots,K-1\},\ \tilk\in\{0,\ldots,\tilK-1\}$\\
    Sample $E = \{\left(\bfe_t^k ,\ldots,  \bfe_{t+H-1}^k\right) \mid \bfe_\tau^k \sim \calN(0, \Sigma),\ k\in\{0,\ldots,\tilK-1\},\ \tau\in\{t,\ldots,t+H\}\}$\\
    Sample $W=\{\left(\bfW_t^{\tilk} ,\ldots,  \bfW_{t+H}^{\tilk}\right)\mid \mathrm{vec}(\bfW^{\tilk}_\tau)\sim\calN(\mathrm{vec}(\bfW^{\text{out}}_{t}), \tilde\Sigma\otimes\bfP_t),\ \tilk\in\{0,\ldots,\tilK-1\},\  \tau\in\{t,\ldots,t+H\}\}$\\

    \For{$k \leftarrow 0$ to $K-1$}{
    $\hat\bfx^k_{t}\leftarrow\bfx_t$\\
    \For{$\tau \leftarrow t$ to $t+H-1$}{
    $\bfv^k_{\tau}\leftarrow\hat\bfu_{\tau} + \bfe_{\tau}^k$\\
    $\hat\bfx^k_{\tau+1} \leftarrow \hat\bfF\left(\hat\bfx^k_{\tau}, \bfv^k_{\tau}\right)$\\
    \For{$\tilk \leftarrow 0$ to $\tilK-1$}{
        $J_{k,\tilk}\pluseq\ell\left(\bfW_\tau^{\tilk\ \top}\hat\bfx^{k}_{\tau}\right) + \lambda(\hat\bfu_{\tau}-\bfu^\text{ref})^{\top} \Sigma^{-1}(\bfv^k_\tau-\bfu^\text{ref}) $\\
    }
    }
    \For{$\tilk \leftarrow 0$ to $\tilK-1$}{
        $J_{k,\tilk} \pluseq\phi\left(\bfW_H^{\tilk\ \top}\hat\bfx^{k}_{H}\right)$\\
    }
    }
    $\tilde\eta \leftarrow \sum_{k=0}^{K-1}\sum_{\tilk=0}^{\tilK-1}\exp\left(-\lambda^{-1} J_{k,\tilk}\right)$\\
    $\hat U^*_t \leftarrow \hat U_t + \tilde\eta^{-1}\sum_{k=0}^{K-1}\left[\sum_{\tilk=0}^{\tilK-1}\exp\left(-\lambda^{-1} J_{k,\tilk}\right) \left(\bfe_t^k ,\ldots,  \bfe_{t+H-1}^k\right)\right]$\\
    \Return{$\hat U^*_t$}
\end{algorithm}

The resulting UMPPI algorithm is presented in~\cref{alg:UMPPI}.
In this algorithm, the expectation in~\cref{eq:importance_sampled_optimal_input} is approximated by Monte Carlo integration using $K$ samples of input noise and $\tilK$ samples of output weight matrices, yielding an approximate optimal input sequence $\hat U^*_t$.
To improve sampling efficiency, we use the input sequence from the previous control cycle, $\hat U_t = \hat U^*_{t-1}$, as the initial estimate of the input sequence $\hat U_t$.
Since the input and output noise are independent random variables, Monte Carlo sampling can be efficiently executed in parallel by pre-generating batches of samples.

\begin{remark}
    Although Proposition \ref{prop:importance_sampled_optimal_input} is derived under the assumption that the admissible input set $\calU$ is unbounded, the algorithm remains practically applicable when $\calU$ is bounded.
    In that case, the updated input sequence can be projected onto $\calU$ after each update step.
    For a detailed discussion of this strategy, see Section III-D2 in Ref.~\cite{Williams2018InformationTheoretic}.
\end{remark}

\subsection{Computational Complexity}\label{sec:complexity}

We analyze the computational complexity of \cref{alg:RPPI}.
The per-step costs are expressed in terms of the reservoir state dimension $\hat{N}$, output dimension $L$, input dimension $M$, prediction horizon $H$, number of input samples $K$, and number of output weight perturbations $\tilK$.

To perform matrix-vector products, the RLS update of the output weight matrix $\bfW_t^{\text{out}}\in\bbR^{\hat{N}\times L}$ and precision matrix $\bfP_t\in\bbR^{\hat{N}\times\hat{N}}$ requires $O(\hat{N}^{2}+L\hat{N})$ operations.
Sampling $K$ Gaussian input noise sequences costs $O(KHM)$.
Drawing $\tilK$ perturbed copies of $\bfW_t^{\text{out}}$ using a Cholesky factorization of $\bfP_t$ requires $O(\hat{N}^{3}+\tilK L\hat{N}^{2})$, where the first term represents the computational cost for preprocessing and the second term represents the computational cost for sampling.
For each trajectory and each prediction step, the ESN performs one reservoir state matrix-vector product, one input matrix-vector product, and $\tilK$ output matrix-vector products, requiring $O(\hat{N}^2+M\hat{N}+\tilK L\hat{N})$ operations per step.
The total trajectory simulation cost is $O(KH(\hat{N}^2+M\hat N+\tilK L\hat N))$.
If the output cost is in quadratic form, evaluating stage and terminal costs $J_{k,\tilk}$ for all $(k,\tilk)$ samples results in $O(KH(M^2+\tilK L^2))$.
Computing weights $\exp(-J_{k,\tilde{k}}/\lambda)$, normalizing factor $\tilde\eta$, and forming the weighted average requires $O(K\tilK+KHM)$.
Hence, the overall complexity per control step $C_{\text{total}}$ is evaluated as follows:
\begin{align}
    \begin{aligned}
        C_{\text{total}} & =
        O(\hat{N}^{2}+L\hat{N})                                                              +O(KHM)     \\
                         & \quad+O(\hat{N}^{3}+\tilK L\hat{N}^{2})                                       \\
                         & \quad+O(KH(\hat{N}^2+M\hat N+\tilK L\hat N))                                  \\
                         & \quad+O(KH(M^2+\tilK L^2)                                      +O(K\tilK+KHM) \\
                         & = O(KH(\hat{N}^2 + M\hat N + \tilK L\hat N + M^2+\tilK L^2)                   \\
                         & \qquad + \hat{N}^{3} + \tilK L\hat N^2).
    \end{aligned}
\end{align}

\begin{remark}\label{rem:complexity}
    Focusing on the computational cost of propagating the dynamics over the prediction horizon while simultaneously evaluating uncertainty, the RPPI framework offers an advantage over probabilistic MPPI variants that propagate uncertainty through the state vector.
    Probabilistic MPPI frameworks that represent uncertainty in the state vector must propagate one copy of the state for every uncertainty sample.
    With state dimension $\hat{N}$, input dimension $M$, and output dimension $L$, a single forward step of the model requires complexity $O(\hat{N}^2+M\hat N+L\hat N)$.
    Sampling $K\times\tilK$ trajectories over a horizon of $H$ therefore costs $O(K\tilK H(\hat{N}^2+M\hat N+L\hat N))$.
    In the proposed UMPPI scheme, the hidden state is shared across all weight samples; only the output map is re-evaluated.
    Therefore, computing $K\times \tilK$ samples requires computational complexity $O(KH(\hat{N}^2+M\hat N+\tilK L\hat N))$.
    For practical settings with $M,L\!\ll\!\hat N$, the dominant computational terms are $O(K\tilK H\hat{N}^2)$ for the former approach and $O(KH\hat N^2 + K\tilK H\hat N)$ for the latter approach.
    Therefore, as $\hat N,H,K,\tilK\to\infty$, the sum of the orders of the UMPPI is smaller than that of the probabilistic MPPI that represent uncertainty in the state vector.
\end{remark}

\section{Numerical Experiments}\label{sec:numerical_experiments}

We validate the proposed methods on two benchmark tasks: position control of a Duffing oscillator and water-level regulation of a four-tank system.
For each task, we compare three methods for computing the control input in \cref{alg:RPPI}:
\begin{enumerate}
    \item QPMPC~\cite{Pan2012Model} (baseline): the input is obtained by minimizing the cost in \cref{eq:cost-function-def}.
          Because the ESN is nonlinear, the model is linearized via a first-order Taylor expansion at every control cycle to yield a QP formulation.
    \item MPPI (proposed RPPI): the input is computed by \cref{alg:UMPPI} with $\tilK=1$ and $\tilde\Sigma=0$.
    \item UMPPI (proposed UMPPI): the input is computed by \cref{alg:UMPPI} with $\tilK\ge1$ and $\tilde\Sigma>0$.
\end{enumerate}

The hyperparameters are determined through preliminary experiments prior to the execution of \cref{alg:RPPI}.
To verify the concept of simultaneous online identification and control, different plant parameters are used during hyperparameter search and the main experiment.
The ESN hyperparameters, namely the sparsity and spectral radius of $\bfW^{\text{res}}$, the leak rate $\alpha$, and the range of the input weight matrix $\bfW^{\text{in}}$, are determined through random search to minimize the cost function \cref{eq:cost-function-RLS} when random input signals are applied.
Furthermore, the MPPI hyperparameter, the inverse temperature $\lambda$, is determined through random search to minimize the cost function \eqref{eq:cost-function-def} when \cref{alg:RPPI} is executed using the ESN model with the optimal hyperparameters.

All numerical simulations were performed on a workstation equipped with an Intel Core i7-12700K processor (12 cores, 20 threads) and 24 GB RAM, running Ubuntu 24.04.1 LTS on Windows Subsystem for Linux 2 (WSL2).
The numerical computation code was implemented in \texttt{Julia 1.10}.
In QPMPC, QP problems were solved with the \texttt{OSQP} package~\cite{osqp}.
At each time step, the ESN model is linearized around the current estimated state $\hat\bfx_t$ to formulate the QP problem.
In RPPI and URPPI, 16-thread parallelization on the CPU was employed to perform parallel sampling of the input sequences.
All source code and data needed to reproduce the results are publicly available at \url{https://github.com/ToyotaCRDL/rppi}.

\subsection{Duffing Oscillator}

We conduct numerical experiments on a Duffing oscillator~\cite{Marconato2012Identification}.
The goal is to control the nonlinear oscillator with cubic stiffness to a reference position using force inputs.
This control system is formulated as
\begin{align}\label{eq:duffing-oscillator}
    \begin{aligned}
        \dot{z}_1(s)      & = z_2(s),                                                                                                        \\
        \dot{z}_2(s)      & = -\frac{k}{m} z_1(s) - \frac{c}{m} z_2(s) - \frac{k_{\mathrm{nl}}}{m} z_1^3(s) + \frac{1}{m} \boldsymbol{u}(s), \\
        \boldsymbol{y}(s) & = z_1(s),
    \end{aligned}
\end{align}
where $s\,[\si{\second}]$ is time,
$[z_1\,z_2]^{\top}\,[\si{\meter},\si{\meter\per\second}]$ are the position and velocity of the oscillator,
$\boldsymbol{u}\,[\si{\newton}]$ is the external force,
and $\boldsymbol{y}=z_1\,[\si{\meter}]$ is the position of the oscillator.
Additionally, $m\,[\si{\kilogram}]$ is the mass,
$k\,[\si{\newton\per\meter}]$ is the linear stiffness coefficient,
$k_{\text{nl}}\,[\si{\newton\per\meter^{3}}]$ is the nonlinear stiffness coefficient,
and $c\,[\si{\newton\second\per\meter}]$ is the damping coefficient.
For such a system, we compute control inputs $\bfu(s)=\bfu_t,\  (s \in [t\Delta s, (t+1)\Delta s), \ t=0,1,\ldots)$ while observing output signals $\bfy_t = \boldsymbol{y}(t\Delta s )\ (t=0,1,\ldots)$ at each control period $\Delta s$~[\si{\second}].
The cost function for determining the control input $U_t=(\bfu_t,\ldots,\bfu_{t+H-1})$ is defined as follows:
\begin{align}
    \begin{aligned}
        J_\text{ctrl}(U_t) & =\sum_{\tau=t}^{t+H-1} \left(\frac{1}{2}Q\,(\bfy_\tau-\boldsymbol{y}^\text{ref})^2 + \frac{1}{2}R\, \bfu_\tau^2\right) \\
                           & \quad +  \frac{1}{2}Q\,(\bfy_{t+H}-\boldsymbol{y}^\text{ref})^2,
    \end{aligned}\label{eq:cost-function-linear-experiment}
\end{align}
where $\boldsymbol{y}^{\text{ref}}\,[\si{\meter}]$ represents the target position of the oscillator,
and $Q>0\,[\si{\meter^{-2}}]$ and $R>0\,[\si{\newton^{-2}}]$ are the weight parameters.

The experimental parameters for the Duffing oscillator system are specified below.
During hyperparameter tuning, the Duffing oscillator was configured with linear stiffness $k = 1.0$, nonlinear stiffness $k_{\text{nl}} = 1.0$, mass $m = 1.0$, and damping $c = 1.0$.
For the main experiments these values were changed to $k = 1.1$, $k_{\text{nl}} = 0.9$, $m = 1.1$, and $c = 0.9$.
The control period was fixed at $\Delta s = 0.1$, and \cref{eq:duffing-oscillator} was integrated with an explicit Euler scheme using this time step.
Cost function weighting employs $Q=100$ for output tracking and $R=1$ for input regulation.
The reference trajectory alternates between positions $\boldsymbol{y}^\text{ref}=1.0$ and $\boldsymbol{y}^\text{ref}=-1.0$ every $20$ seconds.
Control inputs are constrained to the admissible set $\calU=[-10, 10]^{1\times H}$.
The ESN parameters are configured as follows.
The state matrix uses a dimension of $\hat N=400$ with sparsity factor $\rho = 0.48$ (ratio of non-zero connections) and spectral radius of $0.89$.
Input weights $\bfW^\text{in}$ are set to uniform random values in the range $[-2.0, 2.0]$.
The network employs a leak rate of $\alpha=0.59$ and the hyperbolic tangent activation function $f(x)=\tanh(x)$.
For RLS implementation, output weights $\bfW_{-1}^{\text{out}}$ are initialized as uniform random values in $[-0.1, 0.1]$, the precision matrix is set to $\bfP_{-1}=I_{\hat N}$, and the discount factor is $\gamma=1.0$.
The control parameters are specified as follows.
MPPI operates with the input sample size $K=1000$, prediction horizon $H=10$, inverse temperature $\lambda=1.9$, and noise covariance $\Sigma=1.9$.
UMPPI extends these parameters with the output sample size $\tilK=10$ and output weight covariance $\tilde\Sigma=1.0$.

\begin{figure}[t]
    \centering
    \begin{subfigure}{\figwidth}
        \centering
        \includegraphics[width=1\linewidth]{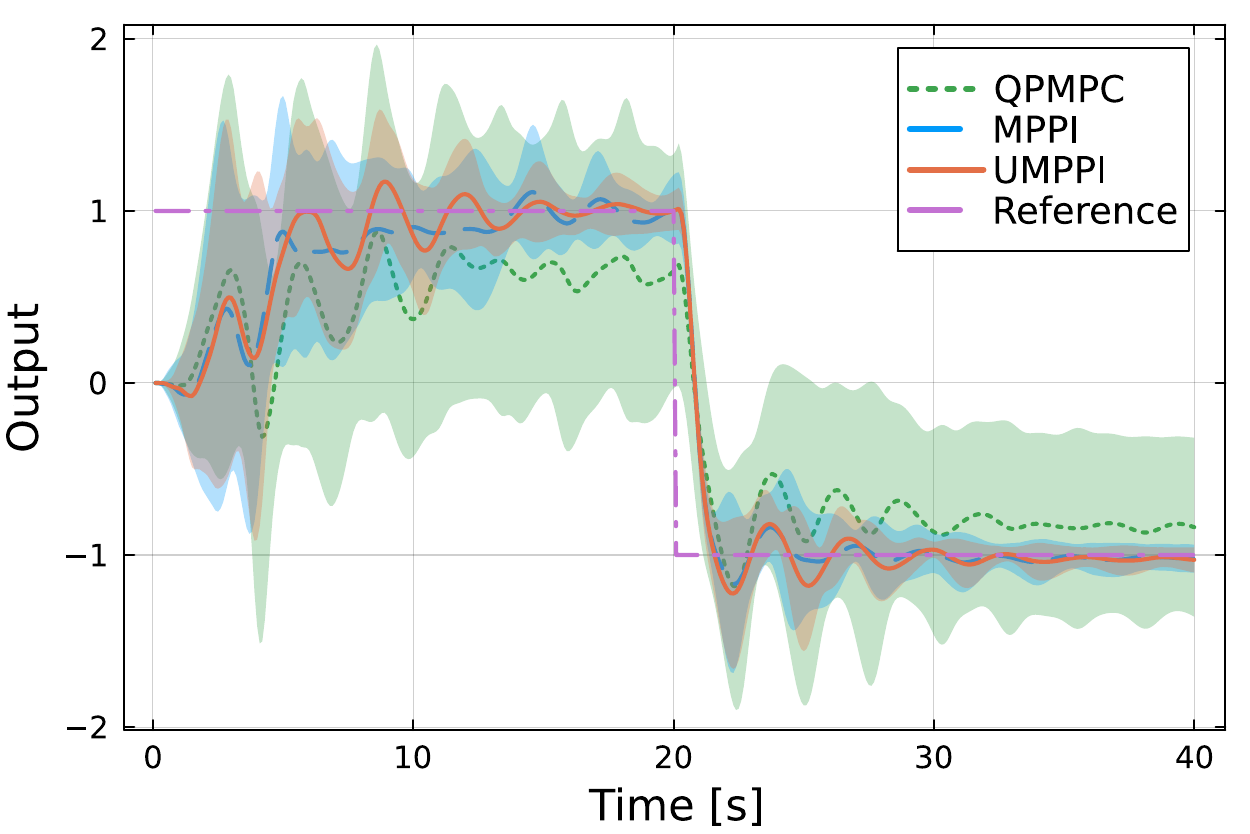}
        \caption{Plant output.}
    \end{subfigure}
    \begin{subfigure}{\figwidth}
        \centering
        \includegraphics[width=1\linewidth]{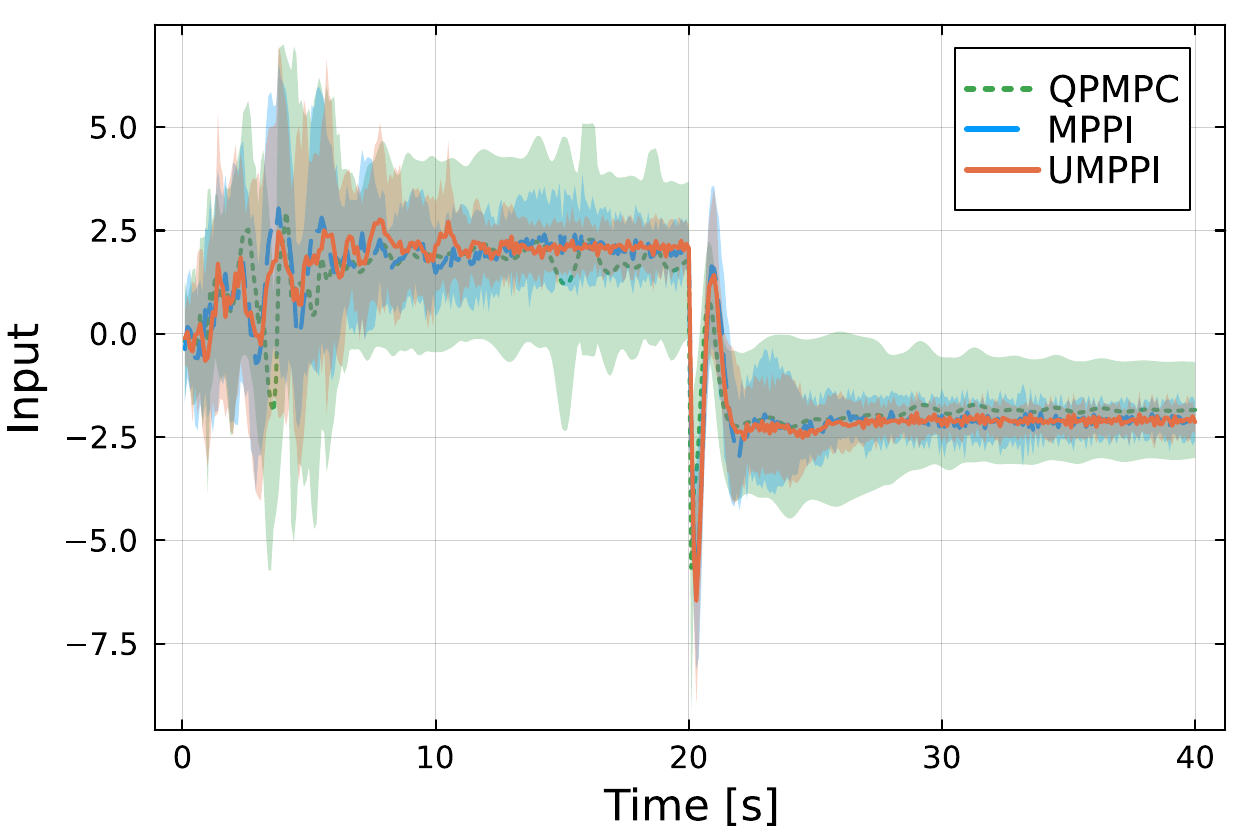}
        \caption{Control input.}
    \end{subfigure}
    \caption{Time response of the Duffing oscillator.
        (a): Time response of the controlled output; (b): Time response of the control input.
        The green dotted line represents QPMPC, the blue dashed line represents MPPI, and the orange solid line represents UMPPI. In (a), the purple dash-dotted line represents the reference output.
        The shaded areas indicate the standard deviation over multiple random-seed runs.
    }
    \label{fig:control_result}
\end{figure}

Figure \ref{fig:control_result} presents the time evolution of system variables under each control method.
Figure \ref{fig:control_result}a displays the output response $\bfy_t$, while \cref{fig:control_result}b shows the corresponding control input $\bfu_t$.
Given the stochastic nature of ESN initialization, MPPI sampling, and UMPPI perturbations, experimental results exhibit inherent variability across executions.
Consequently, experiments were conducted across 20 random seeds; for each seed, we independently resampled (i) the ESN state matrix and input-weight matrix, (ii) the initial output-weight matrix, (iii) the MPPI input-perturbation sequences, and (iv) the UMPPI output-weight perturbation sequences.
Ensemble means are shown as solid lines, and standard deviations are depicted as shaded regions.
The results demonstrate that while QPMPC fails to achieve asymptotic convergence to the reference trajectory, both MPPI and UMPPI successfully attain asymptotic tracking performance.
Moreover, QPMPC exhibits higher variability in both control signals and system outputs compared to the reduced variability observed with MPPI and UMPPI.

\begin{figure}[t]
    \centering
    \includegraphics[width=\figwidth]{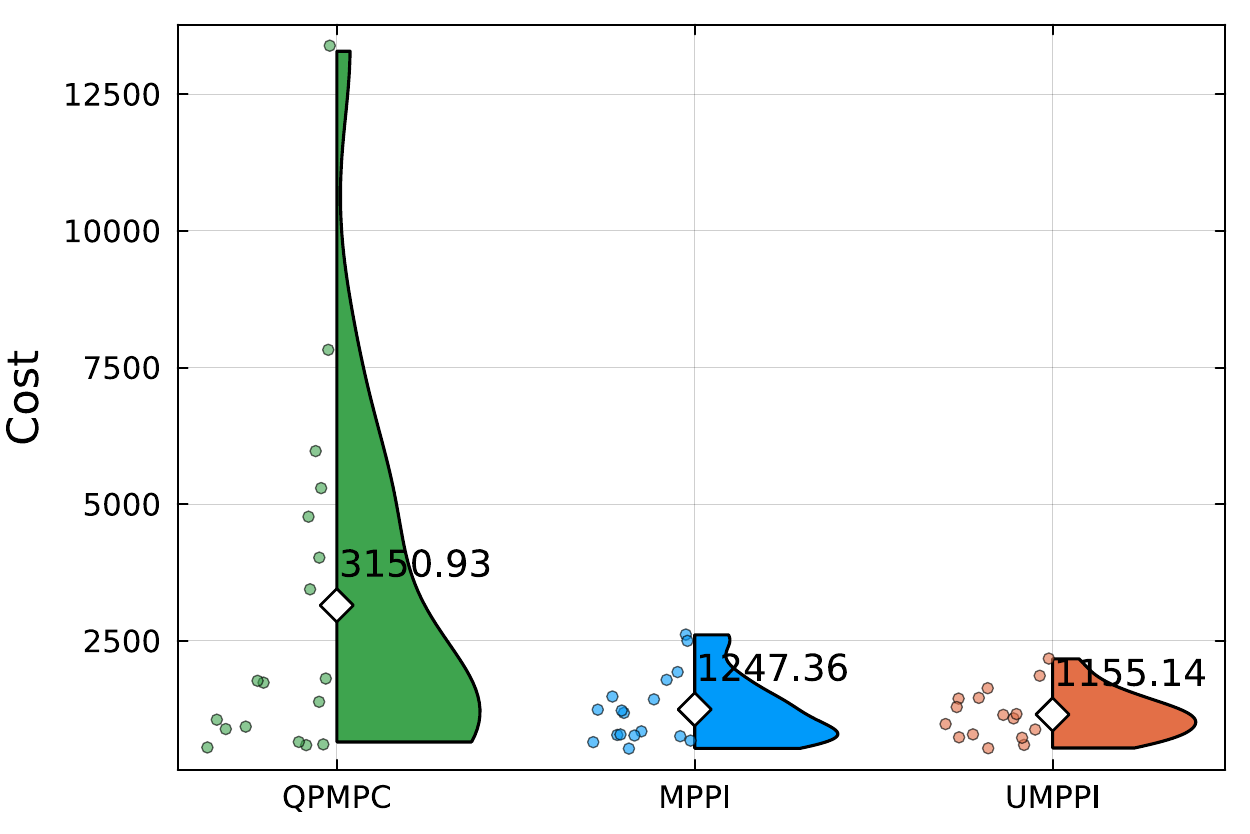}
    \caption{Comparison of the time-integrated control cost in a Duffing oscillator.
        The plot is presented in a raincloud style, combining a violin plot (right) to show distribution density and a scatter plot (left) to display individual data points.
        Diamond markers and numerical values indicate the mean of each distribution.}
    \label{fig:control_cost_comparison}
\end{figure}

\begin{table}[t]
    \centering
    \caption{Comparison of average control cost and computation time among control methods on the Duffing oscillator.}
    \label{tab:cost_time_comparison}
    \begin{tabularx}{0.9\linewidth}{Xcc}
        \toprule
        \textbf{Method}  & \textbf{Control Cost} & \textbf{Computation Time [\si{\second}]} \\
        \midrule
        QPMPC            & 3151                  & 0.008                                    \\
        MPPI (Proposed)  & 1247                  & 0.030                                    \\
        UMPPI (Proposed) & 1155                  & 0.044                                    \\
        \bottomrule
    \end{tabularx}
\end{table}


Figure \ref{fig:control_cost_comparison} presents the distribution of time-integrated control costs across multiple random seeds, specifically measured as
\begin{align}
    \sum_{\tau=0}^{T}
    \left(\frac{1}{2}Q\,(\bfy_\tau-\boldsymbol{y}^\text{ref})^2 + \frac{1}{2}R\, \bfu_\tau^2\right),
\end{align}
where $T$ is the terminal time step, here $T=400$.
Both MPPI and UMPPI achieve substantial reductions in control cost compared to QPMPC, with UMPPI achieving the lowest average cost among the three.
Specifically, UMPPI reduces control cost by approximately 63\% compared to QPMPC and by 7\% compared to MPPI.
This performance improvement can be attributed to UMPPI's ability to account for model uncertainty.
By minimizing the expected cost over the distribution of ESN parameters, UMPPI generates control inputs that are robust to identification errors, avoiding aggressive actions that might be optimal for a nominal model but risky under uncertainty.
As shown in \cref{tab:cost_time_comparison}, MPPI requires 0.030 seconds on average to compute control inputs, while UMPPI requires 0.044 seconds.
Although UMPPI incurs a slight computational overhead due to the additional sampling of output weights, it remains sufficiently fast for online control with a control period of 0.1 seconds.
Considering both control cost and computation time, UMPPI provides improved control performance over MPPI at the expense of only a modest increase in computation time.

\subsection{Four-Tank System}

We conduct numerical experiments on a four-tank system~\cite{Johansson2000quadrupletank}.
The goal is to regulate the water levels in the lower two tanks to their reference values by adjusting the pump flow rates.
This control system is formulated as
\begin{align}\label{eq:fourtank}
    \begin{aligned}
        \dot{z}_{1}(s)    & =
        -\frac{a_{1}}{A_{1}}\sqrt{2g z_{1}(s)}
        +\frac{a_{3}}{A_{1}}\sqrt{2g z_{3}(s)}
        +\frac{b_{1}}{A_{1}} u_{1}(s),                 \\
        \dot{z}_{2}(s)    & =
        -\frac{a_{2}}{A_{2}}\sqrt{2g z_{2}(s)}
        +\frac{a_{4}}{A_{2}}\sqrt{2g z_{4}(s)}
        +\frac{b_{2}}{A_{2}} u_{2}(s),                 \\
        \dot{z}_{3}(s)    & =
        -\frac{a_{3}}{A_{3}}\sqrt{2g z_{3}(s)}
        +\frac{(1-b_{2})}{A_{3}} u_{2}(s),             \\
        \dot{z}_{4}(s)    & =
        -\frac{a_{4}}{A_{4}}\sqrt{2g z_{4}(s)}
        +\frac{(1-b_{1})}{A_{4}} u_{1}(s),             \\
        \boldsymbol{y}(s) & = [z_{1}(s)\ z_2(s)]^\top,
    \end{aligned}
\end{align}
where $z_i\,[\si{\centi\meter}]$ $(i=1,\ldots,4)$ are the water levels in the four tanks and
$\boldsymbol{u}=[u_1\;u_2]^{\top}\,[\si{\centi\meter^{3}\per\second},\si{\centi\meter^{3}\per\second}]$ are the volumetric flow rates delivered by the two pumps.
For $i=1,\ldots,4$,
$A_i\,[\si{\centi\meter^{2}}]$ are the cross-sectional areas of the tanks and
$a_i\,[\si{\centi\meter^{2}}]$ are the outlet areas;
for $i=1,2$, the dimensionless parameters $b_i\in(0,1)$ specify the flow split between upper and lower tanks.
The constant $g\,[\si{\centi\meter\per\second^{2}}]$ denotes gravitational acceleration.
The cost function to determine the control input $U_t=(\bfu_t,\ldots,\bfu_{t+H-1})$ is defined as follows:
\begin{align}
    \begin{aligned}\label{eq:cost-function-linear-experiment_fourtank}
        J_\text{ctrl}(U_t) & = \sum_{\tau=t}^{t+H-1} \left(\frac{1}{2}(\bfy_\tau-\boldsymbol{y}^\text{ref})^\top Q\, (\bfy_\tau-\boldsymbol{y}^\text{ref})\right. \\
                           & \quad +\left.\frac{1}{2}(\bfu_\tau-\boldsymbol{u}^\text{ref})^\top R\, (\bfu_\tau-\boldsymbol{u}^\text{ref})\right)                  \\
                           & \quad + \frac{1}{2}(\bfy_{t+H}-\boldsymbol{y}^\text{ref})^\top Q\, (\bfy_{t+H}-\boldsymbol{y}^\text{ref}),
    \end{aligned}
\end{align}
where $\boldsymbol{y}^{\text{ref}}\,[\si{\centi\meter},\si{\centi\meter}]$ denotes the reference water levels,
$\boldsymbol{u}^{\text{ref}}\,[\si{\centi\meter^{3}\per\second},\si{\centi\meter^{3}\per\second}]$ the reference flow rates, and
$Q,\;R\in\bbR^{2\times2}$ are positive-definite weighting matrices (the entries of $Q$ have units $\si{\centi\meter^{-2}}$,
whereas those of $R$ have units $\si{\second^{2}\centi\meter^{-6}}$).

Experimental parameters for the four-tank system are listed below.
System configuration: tank cross-sectional areas $A_1 = A_3 = 28$, $A_2 = A_4 = 32$; outlet areas $a_1 = a_3 = 0.071$, $a_2 = a_4 = 0.057$; volume-distribution factors were set to $b_1 = 0.7$, $b_2 = 0.6$ during hyperparameter tuning and changed to $b_1 = 0.693$, $b_2 = 0.606$ in the main experiments; sampling period $\Delta s = 1.0$; cost weights $Q = I_2$, $R = I_2$; reference levels alternating between $\boldsymbol{y}^{\text{ref}} = [13\ 13]^\top$ and $\boldsymbol{y}^{\text{ref}} = [11\ 11]^\top$ every 150 s.
ESN configuration: state dimension $\hat N = 400$; reservoir sparsity factor $\rho = 0.17$; spectral radius 0.78; input weights uniformly distributed in $[-1.5, 1.5]$; leak rate $\alpha = 0.078$; activation function $\tanh(x)$.
RLS initialization: output weights $\bfW_{-1}^{\text{out}}$ uniformly distributed in $[-0.1, 0.1]$; precision matrix $\bfP_{-1} = I_{\hat N}$; discount factor $\gamma = 1.0$.
MPPI configuration: reference input $\boldsymbol{u}^{\text{ref}} = [10\ 10]^\top$; input constraints $\calU = [8,\ 12]^{2 \times H}$; sample size $K = 1000$; horizon $H = 10$; inverse temperature $\lambda = 0.71$; noise covariance $\Sigma = 0.71 I_2$.
UMPPI configuration: output sample size $\tilde K = 10$; covariance matrix $\tilde\Sigma = 0.1 I_2$.

\begin{figure}[t]
    \centering
    \begin{subfigure}{\figwidth}
        \centering
        \includegraphics[width=1\linewidth]{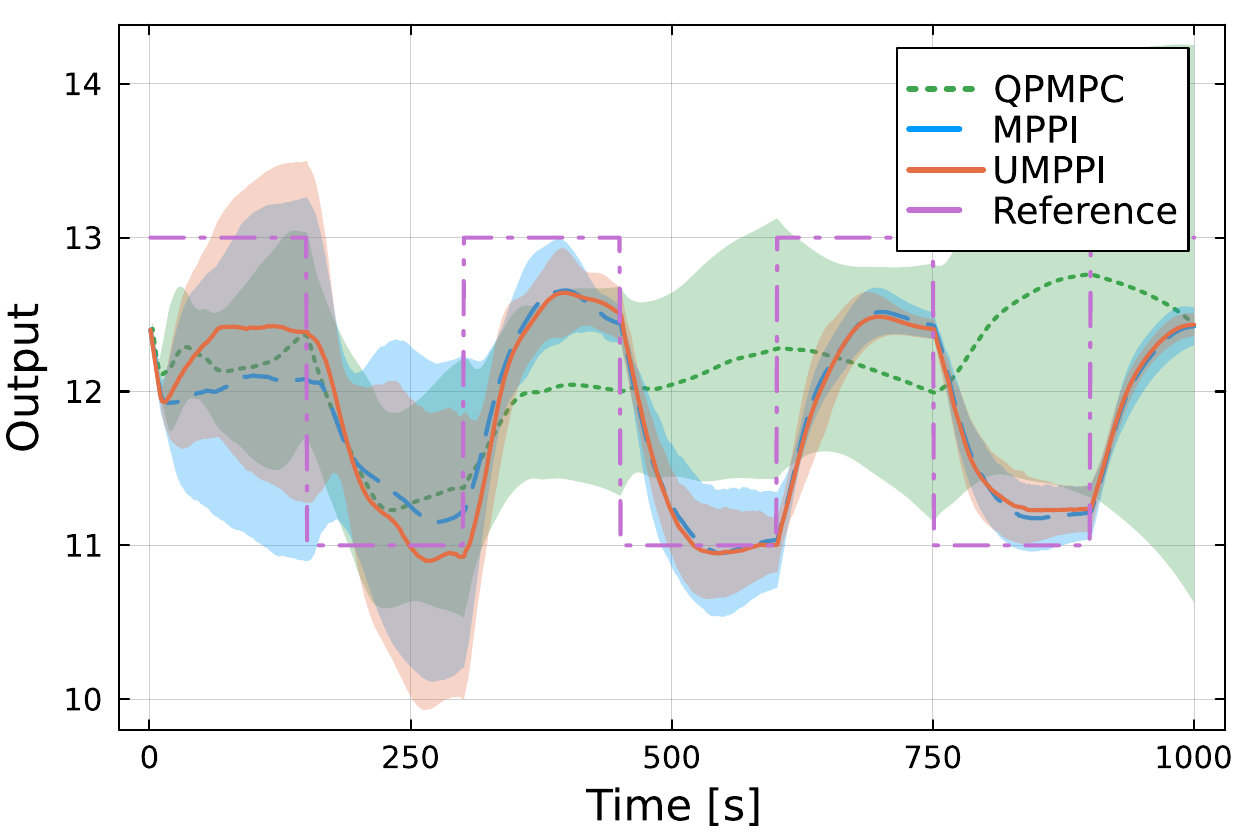}
        \caption{Plant output.}
    \end{subfigure}
    \begin{subfigure}{\figwidth}
        \centering
        \includegraphics[width=1\linewidth]{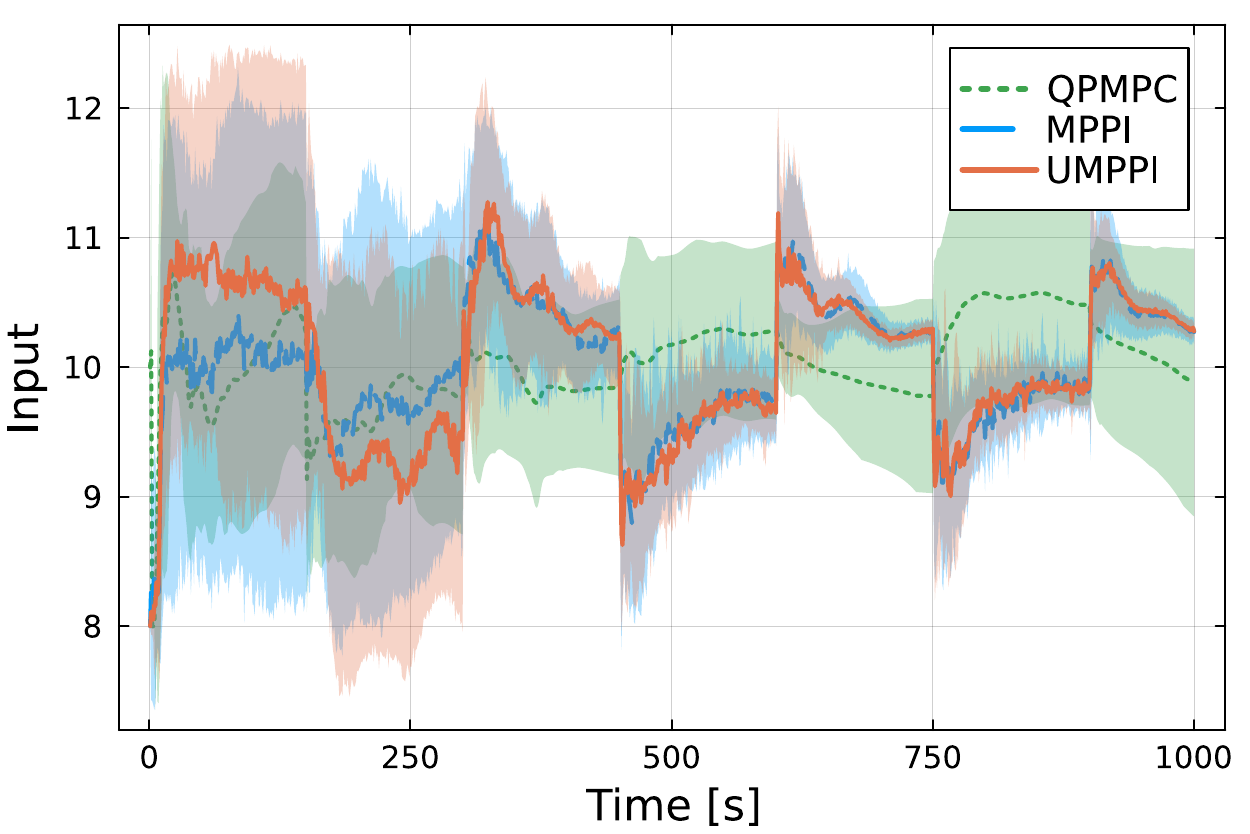}
        \caption{Control input.}
    \end{subfigure}
    \caption{Time response of the four-tank system.
        (a): Time response of the first component of the controlled output in the two-dimensional system;
        (b): Time response of the first component of the control input.
        The green dotted line represents QPMPC, the blue dashed line represents MPPI, and the orange solid line represents UMPPI. In (a), the purple dash-dotted line represents the reference output.
        The shaded areas show the standard deviation over multiple random-seed runs.
    }
    \label{fig:control_result_fourtank}
\end{figure}

The time responses of each variable when controlled by each method are shown in \cref{fig:control_result_fourtank}.
Figure \ref{fig:control_result_fourtank}a shows the time response of the first component of the output $\bfy_t$, and \cref{fig:control_result_fourtank}b shows the time response of the first component of the input $\bfu_t$.
While the plant output under QPMPC gradually deviates from the reference trajectory, both MPPI and UMPPI successfully track the reference whenever it changes.
MPPI and UMPPI can achieve smaller input-output variability compared to QPMPC.

\begin{figure}[t]
    \centering
    \includegraphics[width=\figwidth]{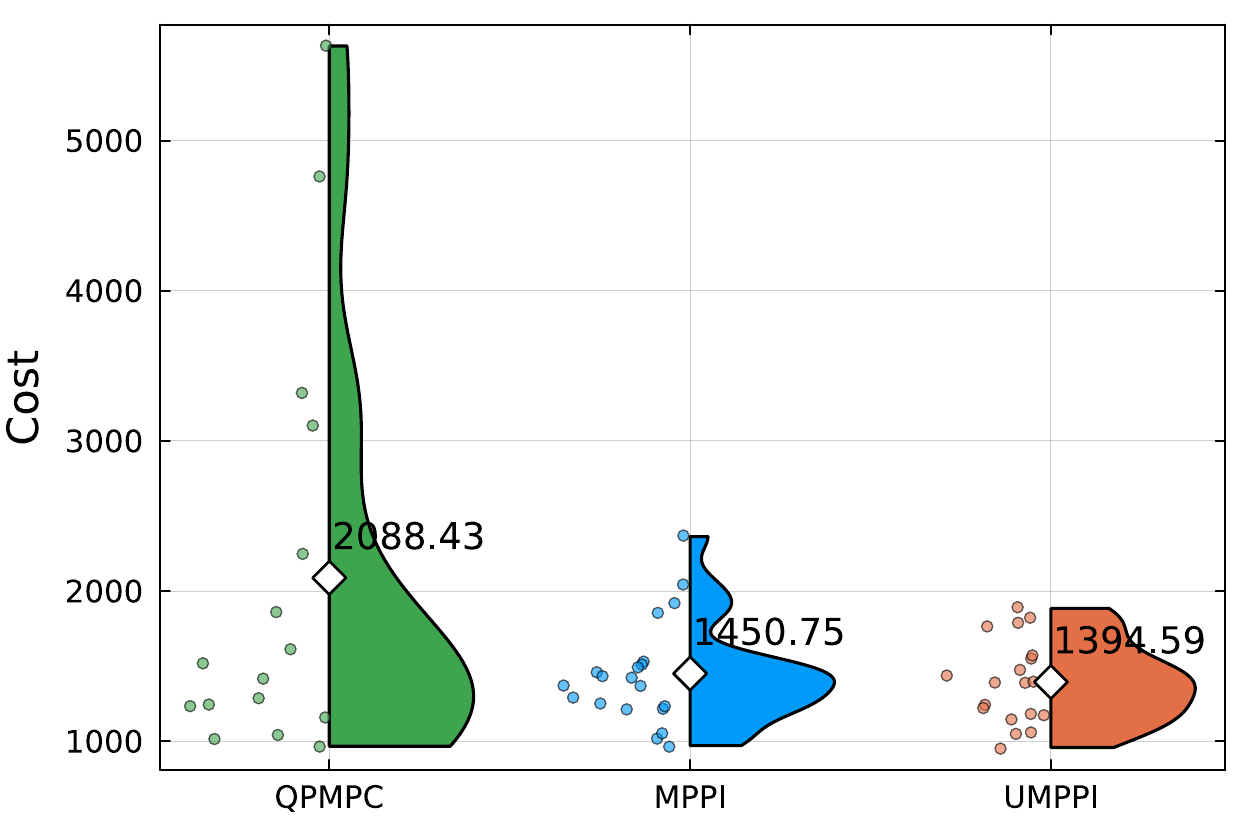}
    \caption{Comparison of the time-integrated control cost in a four-tank system.
        The plot is presented in a raincloud style, combining a violin plot (right) to show distribution density and a scatter plot (left) to display individual data points.
        Diamond markers and numerical values indicate the mean of each distribution.}
    \label{fig:control_cost_comparison_fourtank}
\end{figure}

\begin{table}[t]
    \centering
    \caption{Comparison of average control cost and computation time among control methods on the four-tank system.}
    \label{tab:cost_time_comparison_fourtank}
    \begin{tabularx}{0.9\linewidth}{Xcc}
        \toprule
        \textbf{Method}  & \textbf{Control Cost} & \textbf{Computation Time [\si{\second}]} \\
        \midrule
        QPMPC            & 2088                  & 0.008                                    \\
        MPPI (Proposed)  & 1451                  & 0.033                                    \\
        UMPPI (Proposed) & 1395                  & 0.053                                    \\
        \bottomrule
    \end{tabularx}
\end{table}


The results of computing and visualizing the time-integrated control cost over multiple random seeds, namely
\begin{align}
    \begin{aligned}
        \sum_{\tau=0}^{T} & \left(\frac{1}{2}(\bfy_\tau-\boldsymbol{y}^\text{ref})^\top Q\, (\bfy_\tau-\boldsymbol{y}^\text{ref})\right.    \\
                          & +\left. \frac{1}{2}(\bfu_\tau-\boldsymbol{u}^\text{ref})^\top R\, (\bfu_\tau-\boldsymbol{u}^\text{ref})\right),
    \end{aligned}
\end{align}
are shown in \cref{fig:control_cost_comparison_fourtank}.
Here, the terminal time step $T$ is set to $T=1000$.
Both MPPI and UMPPI achieve substantial reductions in control cost compared to QPMPC, with UMPPI achieving the best overall performance.
Specifically, UMPPI reduces control cost by 33.1\% compared to QPMPC, and by 3.9\% relative to MPPI.
Similar to the Duffing oscillator case, the stochastic formulation of UMPPI allows it to cope with the uncertainty in the learned four-tank dynamics, leading to more robust control performance.
As shown in \cref{tab:cost_time_comparison_fourtank}, MPPI requires 0.033 seconds on average to compute control inputs, while UMPPI requires 0.053 seconds.
Although UMPPI incurs a slight computational overhead due to the additional sampling of output weights, it remains sufficiently fast for online control with a control period of 1.0 seconds.
Overall, UMPPI provides improved control performance over MPPI with only a small additional computational cost.

\section{Conclusion}\label{sec:conclusion}

We presented a control framework that integrates echo state networks (ESN) with model predictive path integral control (MPPI).
Our proposed methods, RPPI and URPPI, combine rapid ESN learning with sampling-based MPPI without linearization to achieve online control for nonlinear systems.
In URPPI, the uncertainty-aware mechanism enables robust control against model identification errors.
Through numerical experiments on the Duffing oscillator and four-tank systems, we demonstrated that URPPI achieves 30-60$\%$ lower control costs than conventional quadratic programming methods, while requiring slightly higher per-step computation times.
Overall, the proposed framework offers a systematic solution for the simultaneous identification and control of nonlinear systems.
Future work includes establishing theoretical stability guarantees for the proposed framework.
Additionally, extending the method to explicitly handle state constraints remains a challenge, as standard MPPI formulations do not naturally enforce hard constraints.
Finally, we aim to pursue hardware validation in robotics, aerospace, and process control applications.

\blind{}{
    \section*{Acknowledgments}
    The authors thank Dr. Hiroaki Yoshida and Ms. Shihori Koyama of Toyota Central R\&D Labs., Inc., and Mr. Takumu Fujioka of Nagoya Institute of Technology, for their helpful discussions and suggestions.
    Generative-AI tools were used to assist with manuscript proofreading and to prototype simulation scripts.
}


\section*{Appendix}

We provide proofs for Propositions \ref{prop:RLS_covariance}--\ref{prop:importance_sampled_optimal_input}.
The results of Proposition \ref{prop:RLS_covariance}
is obtained by extending the analysis in Ref.~\cite{farhang2013adaptive} to the case of ESNs, where the estimated parameters take a matrix form.
The proofs of Propositions \ref{prop:free_energy_inequality}--\ref{prop:importance_sampled_optimal_input} extend the derivation of the MPPI presented in Ref.~\cite{Williams2018InformationTheoretic}.
Specifically, Proposition \ref{prop:free_energy_inequality} parallels Eq.~(13) in Ref.~\cite{Williams2018InformationTheoretic}; Proposition \ref{prop:optimal_distribution} derives along the same route in Section III-A of that work; and Proposition \ref{prop:importance_sampled_optimal_input} corresponds to Eq.~(27) of the same reference.

\begin{refproof}[Proposition \ref{prop:RLS_covariance}]
    From the optimality condition ${\partial J_\text{id}(\bfW^{\text{out}*}_t)}/{\partial \bfW^{\text{out}*}_t} = 0$, we obtain
    \begin{align}\label{eq:optimal_W}
        \bfW^\text{out*}_t = \bfA_t^{-1} \bfB_t.
    \end{align}
    where $\bfB_t \coloneqq \sum_{\tau=0}^t \gamma^{t-\tau} \bfx_\tau \bfy_\tau^\top$.
    From \cref{eq:output_true,eq:optimal_W},
    \begin{align}
        \begin{aligned}
            \bfW^{\text{out}*}_t - \bar \bfW^{\text{out}}
             & = \bfP_t \left(\bfB_t - \bfA_t \bar \bfW^{\text{out}}\right)         \\
             & = \bfP_t \sum_{\tau=0}^t \gamma^{t-\tau} \bfx_\tau \bmeps_\tau^\top.
        \end{aligned}
    \end{align}
    Therefore,
    \begin{align}
        \begin{aligned}
            \mathrm{vec} (\bfW^{\text{out}*}_t - \bar \bfW^{\text{out}})
             & =\begin{bmatrix}
                    \bfP_t\sum_{\tau=0}^t \gamma^{t-\tau} \bfx_\tau  \epsilon_{\tau,1} \\
                    \vdots                                                             \\
                    \bfP_t\sum_{\tau=0}^t \gamma^{t-\tau} \bfx_\tau  \epsilon_{\tau,L}
                \end{bmatrix},
        \end{aligned}
    \end{align}
    where $\epsilon_{\tau,\ell}$ is the $\ell$-th component of $\bmeps_\tau$.
    Using the fact that
    \begin{align}
        \bbE[\epsilon_{k,\ell} \epsilon_{\tilk, \tilde\ell}] =
        \begin{cases}
            \sigma_k^2 & k=\tilk,\ \ell=\tilde \ell, \\
            0          & \text{otherwise},
        \end{cases}
    \end{align}
    we obtain
    \begin{align}
        \begin{aligned}
             & \bbE[\mathrm{vec}(\bfW^{\text{out}*}_t - \bar \bfW^{\text{out}})\mathrm{vec}(\bfW^{\text{out}*}_t - \bar \bfW^{\text{out}})^{\top}] \\
             & = \begin{bmatrix}
                     \sigma_1^2 \bfP_t & 0                 & \cdots & 0                 \\
                     0                 & \sigma_2^2 \bfP_t &        & \vdots            \\
                     \vdots            &                   & \ddots & 0                 \\
                     0                 & \cdots            & 0      & \sigma^2_L \bfP_t
                 \end{bmatrix}                                                                \\
             & =\tilde\Sigma\otimes\bfP_t,
        \end{aligned}
    \end{align}
    where we assumed $\gamma=1$.
    This completes the proof.
\end{refproof}

\begin{refproof}[Proposition \ref{prop:free_energy_inequality}]
    By using Jensen's inequality~\cite{gradshteyn2014table}, the free energy $\calF\left(S, q, \hat\bfx_t, \lambda\right)$ is bounded as
    \begin{align}
         & \calF\left(S, q, \hat\bfx_t, \lambda\right) \label{eq:free-ene-Jensen}                                                        \\
         & \le \bbE_{p_{U_t}}\left[S\left(V_t,W_t ; \hat\bfx_t\right)+ \lambda  \log\frac{p_{U_t}(V_t,W_t)}{q(V_t,W_t)}\right].\nonumber
    \end{align}
    For the second term, by taking the reference probability density function $q$ as \cref{eq:base_prob_V,eq:base_prob_W}, it follows
    \begin{align}
        \begin{aligned}  \label{eq:KL}
             & \bbE_{p_{U_t}}\left[\log\frac{p_{U_t}(V_t,W_t)}{q(V_t,W_t)}\right]                                           \\
             & = \bbE_{p_{U_t}}\left[\log\frac{p(V_t|U_t,\Sigma)}{q(V_t)}\right]                                            \\
             & =\frac{1}{2} \sum_{\tau=t}^{t+H-1}(\bfu_\tau-\bfu^\text{ref})^{\top} \Sigma^{-1}(\bfu_\tau-\bfu^\text{ref}).
        \end{aligned}
    \end{align}
    Substituting \cref{eq:KL} into \cref{eq:free-ene-Jensen} completes the proof.
\end{refproof}

\begin{refproof}[Proposition \ref{prop:optimal_distribution}]
    In \cref{eq:free-ene-Jensen}, taking the probability density function of $V_t$ and $W_t$ as \cref{eq:optimal_distribution}, we obtain
    \begin{align}
        \begin{aligned}\label{eq:free_energy_1st}
             & \bbE_{p^*}\left[\hat J_\text{ctrl}(V_t,W_t)\right]                                                         \\
             & = \bbE_{p^*}\left[S\left(V_t,W_t ; \hat\bfx_t\right)+ \lambda  \log\frac{p^*(V_t,W_t)}{q(V_t,W_t)}\right],
        \end{aligned}
    \end{align}
    For the second term on the right-hand side,
    \begin{align}
        \begin{aligned}\label{eq:free_energy_2nd}
             & \bbE_{p^*}\left[\log\frac{p^*(V_t,W_t)}{q(V_t,W_t)}\right]                                                                       \\
             & = \bbE_{p^*}\left[\log\left(\frac{1}{\eta} \exp \left(-\frac{1}{\lambda} S\left(V_t,W_t ; \hat\bfx_t\right)\right)\right)\right] \\
             & = -\frac{1}{\lambda}\bbE_{p^*}\left[S\left(V_t,W_t ; \hat\bfx_t\right)\right] - \log(\eta).
        \end{aligned}
    \end{align}
    By substituting \cref{eq:free_energy_2nd} into \cref{eq:free_energy_1st},
    \begin{align}
        \begin{aligned}
             & \bbE_{p^*}\left[\hat J_\text{ctrl}(V_t,W_t)\right]  = -\lambda \log(\eta)
            = \calF\left(S, q, \hat\bfx_t, \lambda\right).
        \end{aligned}
    \end{align}
    This completes the proof.
\end{refproof}

\begin{refproof}[Proposition \ref{prop:importance_sampled_optimal_input}]
    First, we show that the optimal input satisfies $\bfu_\tau^{*}=\bbE_{p^*}[\bfv_\tau]$.
    This is shown as follows:
    \begin{align}
        U_t^{*}
         & =\underset{U_t \in \calU}{\argmin}\left\{\bbE_{p^*}\left[\log \left(\frac{p^*(V_t,W_t)}{p_{U_t}(V_t,W_t)}\right)\right]\right\}   \nonumber                                                  \\
         & =\underset{U_t \in \calU}{\argmax}\left\{\bbE_{p^*}[\log (p(V_t | U_t, \Sigma))]\right\}                                                                                                     \\
         & = \underset{U_t \in \calU}{\argmin} \left\{ \bbE_{p^*}\left[\sum_{\tau=0}^{H-1}\left(\bfv_\tau-\bfu_\tau\right)^{\top} \Sigma^{-1}\left(\bfv_\tau-\bfu_\tau\right)\right] \right\}.\nonumber
    \end{align}
    In the third line, $p(V_t | U_t, \Sigma)$ defined in \cref{eq:prob_V} was substituted.
    When $\calU=\bbR^{M\times H}$, the solution to this optimization problem satisfies $\bfu_\tau^{*}=\bbE_{p^*}[\bfv_\tau],\ (\tau=t,\ldots,t+H-1)$.

    Next, by introducing importance sampling using $\hat U_t$, we show \cref{eq:importance_sampled_optimal_input}.
    The above optimal input is expressed as
    \begin{align}
        \begin{aligned}
             & \bbE_{p^*}[\bfv_\tau]= \int \int  \bfv_\tau  p^*(V_t,W_t) \rmd V_t\rmd W_t                                   \\
             & = \int \int  \bfv_\tau  \frac{p^*(V_t,W_t)}{p_{\hat U_t}(V_t,W_t)} \, p_{\hat U_t}(V_t,W_t) \rmd V_t\rmd W_t \\
             & = \bbE_{p_{\hat U_t}}[w^{\prime}_t(V_t,W_t) \bfv_\tau],
        \end{aligned}\label{eq:expected_value}
    \end{align}
    where
    \begin{align}\label{eq:w_t}
        w^{\prime}_t(V_t,W_t) & \coloneqq \frac{p^*(V_t,W_t)}{p_{\hat U_t}(V_t,W_t)}.
    \end{align}
    By the definition of $p^*$ and $p_{\hat U_t}$ in \cref{eq:optimal_distribution,eq:prob_U},
    \begin{align}\label{eq:w_t_modified}
        w^{\prime}_t(V_t,W_t) & =\frac{\frac{1}{\eta} \exp \left(-\frac{1}{\lambda} S\left(V_t,W_t ; \hat\bfx_t\right)\right) q(V_t)}{p(V_t|\hat U_t, \Sigma)}.
    \end{align}
    Using the reference density function $q(V_t)$ in \cref{eq:base_prob_V}, we have
    \begin{align}
        \begin{aligned}
              & \frac{q(V_t)}{p(V_t|\hat U_t, \Sigma)}
            =  \frac{\prod_{\tau=t}^{t+H-1} \calN(\bfv_\tau|\bfu^\text{ref},\Sigma)}{\prod_{\tau=t}^{t+H-1} \calN(\bfv_\tau|\hat\bfu_\tau,\Sigma)} \\
            = & \exp\left(-\sum_{\tau=t}^{t+H-1}(\hat\bfu_\tau-\bfu^\text{ref})^\top\Sigma^{-1}(\bfv_\tau-\bfu^\text{ref})\right)                  \\
              & \cdot \exp\left(\frac12\sum_{\tau=t}^{t+H-1}(\hat\bfu_\tau-\bfu^\text{ref})^\top\Sigma^{-1}(\hat\bfu_\tau-\bfu^\text{ref})\right). \\
        \end{aligned}\label{eq:q/p}
    \end{align}
    Substituting this into \eqref{eq:w_t_modified}, we obtain
    \begin{align}\label{eq:w_t_modified_substituted}
        w^{\prime}_t(V_t,W_t) = \frac{1}{\eta^{\prime}} w_t(V_t,W_t),
    \end{align}
    where
    \begin{align}
        \eta^{\prime} \coloneqq \exp\left(-\frac12\sum_{\tau=t}^{t+H-1}(\hat\bfu_\tau-\bfu^\text{ref})^\top\Sigma^{-1}(\hat\bfu_\tau-\bfu^\text{ref})\right)  \eta.
    \end{align}
    Here, $\eta^{\prime}$ can be transformed as follows:
    \begin{align}
        \begin{aligned}
            \eta^{\prime} & = \bbE_{p^*}\left[\eta^{\prime}\right]                                                                                                                   \\
                          & =\bbE_{p_{\hat U_t}}\left[\eta^{\prime}\ \frac{p^*(V_t,W_t)}{p(V_t|\hat U_t, \Sigma) p(W_t|\bfW_t^{\text{out}},\bfP_t,\tilde\Sigma)}\right]              \\
                          & = \bbE_{p_{\hat U_t}}\left[\eta^{\prime}\ w^{\prime}_t(V_t,W_t) \right]                                                                                  \\
                          & =\bbE_{p_{\hat U_t}}\left[w_t(V_t,W_t) \right]                                                                                             =\tilde \eta.
        \end{aligned}\label{eq:tilde_eta}
    \end{align}
    Substituting \cref{eq:w_t_modified_substituted,eq:tilde_eta} into \cref{eq:expected_value} completes the proof.
\end{refproof}

\vfill


\begin{thebibliography}{10}
    \providecommand{\url}[1]{#1}
    \csname url@samestyle\endcsname
    \providecommand{\newblock}{\relax}
    \providecommand{\bibinfo}[2]{#2}
    \providecommand{\BIBentrySTDinterwordspacing}{\spaceskip=0pt\relax}
    \providecommand{\BIBentryALTinterwordstretchfactor}{4}
    \providecommand{\BIBentryALTinterwordspacing}{\spaceskip=\fontdimen2\font plus
        \BIBentryALTinterwordstretchfactor\fontdimen3\font minus \fontdimen4\font\relax}
    \providecommand{\BIBforeignlanguage}[2]{{%
                \expandafter\ifx\csname l@#1\endcsname\relax
                    \typeout{** WARNING: IEEEtran.bst: No hyphenation pattern has been}%
                    \typeout{** loaded for the language `#1'. Using the pattern for}%
                    \typeout{** the default language instead.}%
                \else
                    \language=\csname l@#1\endcsname
                \fi
                #2}}
    \providecommand{\BIBdecl}{\relax}
    \BIBdecl

    \bibitem{Ljung1999System}
    L.~Ljung, \emph{System Identification (2nd Ed.): Theory for the User}.\hskip 1em plus 0.5em minus 0.4em\relax USA: Prentice Hall PTR, 1999.

    \bibitem{Lewis2012Optimal}
    F.~L. Lewis, D.~Vrabie, and V.~L. Syrmos, \emph{Optimal Control}.\hskip 1em plus 0.5em minus 0.4em\relax John Wiley \& Sons, 2012.

    \bibitem{nijmeijer1990nonlinear}
    H.~Nijmeijer and A.~{Van der Schaft}, \emph{Nonlinear Dynamical Control Systems}.\hskip 1em plus 0.5em minus 0.4em\relax Springer, 1990, vol. 175.

    \bibitem{Pan2008Nonlinear}
    Y.~Pan and J.~Wang, ``Nonlinear {{Model Predictive Control Using}} a {{Recurrent Neural Network}},'' in \emph{2008 {{IEEE International Joint Conference}} on {{Neural Networks}} ({{IEEE World Congress}} on {{Computational Intelligence}})}, Jun. 2008, pp. 2296--2301.

    \bibitem{Chen2018Optimala}
    Y.~Chen, Y.~Shi, and B.~Zhang, ``Optimal {{Control Via Neural Networks}}: {{A Convex Approach}},'' in \emph{International {{Conference}} on {{Learning Representations}}}, Sep. 2018.

    \bibitem{Yan2012Model}
    Z.~Yan and J.~Wang, ``Model {{Predictive Control}} of {{Nonlinear Systems With Unmodeled Dynamics Based}} on {{Feedforward}} and {{Recurrent Neural Networks}},'' \emph{IEEE Transactions on Industrial Informatics}, vol.~8, no.~4, pp. 746--756, Jan. 2012.

    \bibitem{Bonassi2021Nonlinear}
    F.~Bonassi, C.~F.~O. {da Silva}, and R.~Scattolini, ``Nonlinear {{MPC}} for {{Offset-Free Tracking}} of systems learned by {{GRU Neural Networks}},'' \emph{IFAC-PapersOnLine}, vol.~54, no.~14, pp. 54--59, Jan. 2021.

    \bibitem{Terzi2021Learning}
    E.~Terzi, F.~Bonassi, M.~Farina, and R.~Scattolini, ``Learning {{Model Predictive Control}} with {{Long Short-Term Memory Networks}},'' \emph{International Journal of Robust and Nonlinear Control}, vol.~31, no.~18, pp. 8877--8896, 2021.

    \bibitem{Cao2017Gaussiana}
    G.~Cao, E.~M.-K. Lai, and F.~Alam, ``Gaussian {{Process Model Predictive Control}} of {{Unknown Non-linear Systems}},'' \emph{IET Control Theory \& Applications}, vol.~11, no.~5, pp. 703--713, Mar. 2017.

    \bibitem{Polcz2023Efficient}
    P.~Polcz, T.~P{\'e}ni, and R.~T{\'o}th, ``Efficient {{Implementation}} of {{Gaussian Process}}--based {{Predictive Control}} by {{Quadratic Programming}},'' \emph{IET Control Theory \& Applications}, vol.~17, no.~8, pp. 968--984, 2023.

    \bibitem{Salzmann2023RealTime}
    T.~Salzmann, E.~Kaufmann, J.~Arrizabalaga, M.~Pavone, D.~Scaramuzza, and M.~Ryll, ``Real-{{Time Neural MPC}}: {{Deep Learning Model Predictive Control}} for {{Quadrotors}} and {{Agile Robotic Platforms}},'' \emph{IEEE Robotics and Automation Letters}, vol.~8, no.~4, pp. 2397--2404, Apr. 2023.

    \bibitem{williams2006gaussian}
    C.~K. Williams and C.~E. Rasmussen, \emph{Gaussian Processes for Machine Learning}.\hskip 1em plus 0.5em minus 0.4em\relax MIT Press Cambridge, MA, 2006, vol.~2.

    \bibitem{Park2020Gaussian}
    J.~Park and J.~Choi, ``Gaussian {{Process Online Learning With}} a {{Sparse Data Stream}},'' \emph{IEEE Robotics and Automation Letters}, vol.~5, no.~4, pp. 5977--5984, Oct. 2020.

    \bibitem{Alessio2009Survey}
    A.~Alessio and A.~Bemporad, ``A {{Survey}} on {{Explicit Model Predictive Control}},'' in \emph{Nonlinear {{Model Predictive Control}}: {{Towards New Challenging Applications}}}, ser. Lecture {{Notes}} in {{Control}} and {{Information Sciences}}.\hskip 1em plus 0.5em minus 0.4em\relax Berlin, Heidelberg: Springer, 2009, pp. 345--369.

    \bibitem{Kvasnica2013Complexity}
    M.~Kvasnica, J.~Hled{\'i}k, I.~Rauov{\'a}, and M.~Fikar, ``Complexity {{Reduction}} of {{Explicit Model Predictive Control}} via {{Separation}},'' \emph{Automatica}, vol.~49, no.~6, pp. 1776--1781, Jun. 2013.

    \bibitem{Chen2018Approximating}
    S.~Chen, K.~Saulnier, N.~Atanasov, D.~D. Lee, V.~Kumar, G.~J. Pappas, and M.~Morari, ``Approximating {{Explicit Model Predictive Control Using Constrained Neural Networks}},'' in \emph{2018 {{Annual American Control Conference}} ({{ACC}})}, Jun. 2018, pp. 1520--1527.

    \bibitem{Forssell1999Closedloop}
    U.~Forssell and L.~Ljung, ``Closed-{{Loop Identification Revisited}},'' \emph{Automatica}, vol.~35, no.~7, pp. 1215--1241, Jul. 1999.

    \bibitem{Qin2006overview}
    S.~J. Qin, ``An {{Overview}} of {{Subspace Identification}},'' \emph{Computers \& Chemical Engineering}, vol.~30, no.~10, pp. 1502--1513, Sep. 2006.

    \bibitem{Jaeger2002Adaptive}
    H.~Jaeger, ``Adaptive {{Nonlinear System Identification}} with {{Echo State Networks}},'' in \emph{Advances in {{Neural Information Processing Systems}}}, vol.~15.\hskip 1em plus 0.5em minus 0.4em\relax MIT Press, 2002.

    \bibitem{Jaeger2007Optimization}
    H.~Jaeger, M.~Luko{\v s}evi{\v c}ius, D.~Popovici, and U.~Siewert, ``Optimization and {{Applications}} of {{Echo State Networks}} with {{Leaky-Integrator Neurons}},'' \emph{Neural Networks}, vol.~20, no.~3, pp. 335--352, Apr. 2007.

    \bibitem{Salmen2005Echo}
    M.~Salmen and P.~G. Ploger, ``Echo {{State Networks}} used for {{Motor Control}},'' in \emph{Proceedings of the 2005 {{IEEE International Conference}} on {{Robotics}} and {{Automation}}}, Apr. 2005, pp. 1953--1958.

    \bibitem{Sun2024Systematic}
    C.~Sun, M.~Song, D.~Cai, B.~Zhang, S.~Hong, and H.~Li, ``A {{Systematic Review}} of {{Echo State Networks From Design}} to {{Application}},'' \emph{IEEE Transactions on Artificial Intelligence}, vol.~5, no.~1, pp. 23--37, Jan. 2024.

    \bibitem{Pan2012Model}
    Y.~Pan and J.~Wang, ``Model {{Predictive Control}} of {{Unknown Nonlinear Dynamical Systems Based}} on {{Recurrent Neural Networks}},'' \emph{IEEE Transactions on Industrial Electronics}, vol.~59, no.~8, pp. 3089--3101, Aug. 2012.

    \bibitem{Jordanou2022Echo}
    J.~P. Jordanou, E.~A. Antonelo, and E.~Camponogara, ``Echo {{State Networks}} for {{Practical Nonlinear Model Predictive Control}} of {{Unknown Dynamic Systems}},'' \emph{IEEE Transactions on Neural Networks and Learning Systems}, vol.~33, no.~6, pp. 2615--2629, Jun. 2022.

    \bibitem{Schwedersky2022Adaptive}
    B.~B. Schwedersky, R.~C.~C. Flesch, and S.~B. Rovea, ``Adaptive {{Practical Nonlinear Model Predictive Control}} for {{Echo State Network Models}},'' \emph{IEEE Transactions on Neural Networks and Learning Systems}, vol.~33, no.~6, pp. 2605--2614, Jun. 2022.

    \bibitem{Schwedersky2022Echo}
    ------, ``Echo {{State Networks}} for {{Online}}, {{Multi-step MPC Relevant Identification}},'' \emph{Engineering Applications of Artificial Intelligence}, vol. 108, p. 104596, Feb. 2022.

    \bibitem{Williams2018InformationTheoretic}
    G.~Williams, P.~Drews, B.~Goldfain, J.~M. Rehg, and E.~A. Theodorou, ``Information-{{Theoretic Model Predictive Control}}: {{Theory}} and {{Applications}} to {{Autonomous Driving}},'' \emph{IEEE Transactions on Robotics}, vol.~34, no.~6, pp. 1603--1622, Feb. 2018.

    \bibitem{Williams2015Model}
    G.~Williams, A.~Aldrich, and E.~Theodorou, ``Model {{Predictive Path Integral Control}} using {{Covariance Variable Importance Sampling}},'' \emph{arXiv:1509.01149 [cs]}, Oct. 2015.

    \bibitem{Kazim2024Recent}
    M.~Kazim, J.~Hong, M.-G. Kim, and K.-K.~K. Kim, ``Recent {{Advances}} in {{Path Integral Control}} for {{Trajectory Optimization}}: {{An Overview}} in {{Theoretical}} and {{Algorithmic Perspectives}},'' \emph{Annual Reviews in Control}, vol.~57, p. 100931, Jan. 2024.

    \bibitem{Sanchez-Gonzalez2018Graph}
    A.~{Sanchez-Gonzalez}, N.~Heess, J.~T. Springenberg, J.~Merel, M.~Riedmiller, R.~Hadsell, and P.~Battaglia, ``Graph {{Networks}} as {{Learnable Physics Engines}} for {{Inference}} and {{Control}},'' in \emph{Proceedings of the 35th {{International Conference}} on {{Machine Learning}}}.\hskip 1em plus 0.5em minus 0.4em\relax PMLR, Jul. 2018, pp. 4470--4479.

    \bibitem{Park2023Simultaneous}
    J.~Park, M.~R. Babaei, S.~A. Munoz, A.~N. Venkat, and J.~D. Hedengren, ``Simultaneous {{Multistep Transformer Architecture}} for {{Model Predictive Control}},'' \emph{Computers \& Chemical Engineering}, vol. 178, p. 108396, Oct. 2023.

    \bibitem{Wang2024Fast}
    W.~Wang, H.~Zhang, Y.~Wang, Y.~Tian, and Z.~Wu, ``Fast {{Explicit Machine Learning-Based Model Predictive Control}} of {{Nonlinear Processes Using Input Convex Neural Networks}},'' \emph{Industrial \& Engineering Chemistry Research}, vol.~63, no.~40, pp. 17\,279--17\,293, Oct. 2024.

    \bibitem{MoerlandModelbased2023}
    T.~M. Moerland, J.~Broekens, A.~Plaat, and C.~M. Jonker, ``Model-based {{Reinforcement Learning}}: {{A Survey}},'' \emph{Foundations and Trends in Machine Learning}, vol.~16, no.~1, pp. 1--118, Jan. 2023.

    \bibitem{Brunke2022Safe}
    L.~Brunke, M.~Greeff, A.~W. Hall, Z.~Yuan, S.~Zhou, J.~Panerati, and A.~P. Schoellig, ``Safe {{Learning}} in {{Robotics}}: {{From Learning-Based Control}} to {{Safe Reinforcement Learning}},'' \emph{Annual Review of Control, Robotics, and Autonomous Systems}, vol.~5, no. Volume 5, 2022, pp. 411--444, May 2022.

    \bibitem{Chua2018Deep}
    K.~Chua, R.~Calandra, R.~McAllister, and S.~Levine, ``Deep {{Reinforcement Learning}} in a {{Handful}} of {{Trials}} using {{Probabilistic Dynamics Models}},'' in \emph{Advances in {{Neural Information Processing Systems}}}, vol.~31.\hskip 1em plus 0.5em minus 0.4em\relax Curran Associates, Inc., 2018.

    \bibitem{Pozzi2025Imitation}
    A.~Pozzi, A.~Incremona, and D.~Toti, ``Imitation {{Learning-driven Approximation}} of {{Stochastic Control Models}},'' \emph{Applied Intelligence}, vol.~55, no.~12, p. 838, Jul. 2025.

    \bibitem{williams2018robust}
    G.~Williams, B.~Goldfain, P.~Drews, K.~Saigol, J.~M. Rehg, and E.~A. Theodorou, ``Robust {{Sampling Based Model Predictive Control}} with {{Sparse Objective Information}}.'' in \emph{Robotics: {{Science}} and Systems}, vol.~14, 2018, p. 2018.

    \bibitem{Yin2023RiskAware}
    J.~Yin, Z.~Zhang, and P.~Tsiotras, ``Risk-{{Aware Model Predictive Path Integral Control Using Conditional Value-at-Risk}},'' in \emph{2023 {{IEEE International Conference}} on {{Robotics}} and {{Automation}} ({{ICRA}})}, May 2023, pp. 7937--7943.

    \bibitem{Mesbah2018Stochastic}
    A.~Mesbah, ``Stochastic {{Model Predictive Control}} with {{Active Uncertainty Learning}}: {{A Survey}} on {{Dual Control}},'' \emph{Annual Reviews in Control}, vol.~45, pp. 107--117, Jan. 2018.

    \bibitem{farhang2013adaptive}
    B.~{Farhang-Boroujeny}, \emph{Adaptive Filters: Theory and Applications}.\hskip 1em plus 0.5em minus 0.4em\relax John Wiley \& Sons, 2013.

    \bibitem{haykin2002adaptive}
    S.~S. Haykin, \emph{Adaptive Filter Theory}.\hskip 1em plus 0.5em minus 0.4em\relax Pearson Education India, 2002.

    \bibitem{osqp}
    B.~Stellato, G.~Banjac, P.~Goulart, A.~Bemporad, and S.~Boyd, ``{{OSQP}}: {{An Operator Splitting Solver}} for {{Quadratic Programs}},'' \emph{Mathematical Programming Computation}, vol.~12, no.~4, pp. 637--672, 2020.

    \bibitem{Marconato2012Identification}
    A.~Marconato, J.~Sj{\"o}berg, J.~Suykens, and J.~Schoukens, ``Identification of the {{Silverbox Benchmark Using Nonlinear State-Space Models}},'' \emph{IFAC Proceedings Volumes}, vol.~45, no.~16, pp. 632--637, Jul. 2012.

    \bibitem{Johansson2000quadrupletank}
    K.~Johansson, ``The {{Quadruple-Tank Process}}: {{A Multivariable Laboratory Process}} with an {{Adjustable Zero}},'' \emph{IEEE Transactions on Control Systems Technology}, vol.~8, no.~3, pp. 456--465, May 2000.

    \bibitem{gradshteyn2014table}
    I.~S. Gradshteyn and I.~M. Ryzhik, \emph{Table of Integrals, Series, and Products}.\hskip 1em plus 0.5em minus 0.4em\relax Academic Press, 2014.

\end{thebibliography}
\end{document}